\documentclass[aps,amssymb,amsmath,amsfonts,pra,superscriptaddress, twocolumn, a4paper, accepted=2020-01-27]{quantumarticle}
\pdfoutput=1
\usepackage[utf8]{inputenc}
\usepackage{amsmath}
\usepackage{graphicx}
\usepackage{amssymb} 
\usepackage{qcircuit}
\usepackage{xy}
\usepackage{supertabular}
\usepackage{tabularx}
\usepackage{subfigure}
\usepackage{soul}
\usepackage{multirow}

\usepackage[english]{babel}
\def\la{{\langle}}
\def\ra{{\rangle}}
\newcommand{\ket}[1]{|#1\ra}
\newcommand{\bra}[1]{\la #1|}
\newcommand{\braket}[2]{\la #1|#2 \ra}

\usepackage{color}

\begin{document}
	
\title{Data re-uploading for a universal quantum classifier}

\author{Adrián P\'{e}rez-Salinas}
\affiliation{Barcelona Supercomputing Center}
\affiliation{Institut de Ci\`encies del Cosmos, Universitat de Barcelona, Barcelona, Spain}
\orcid{0000-0001-5430-6468}
\author{Alba Cervera-Lierta}
\affiliation{Barcelona Supercomputing Center}
\affiliation{Institut de Ci\`encies del Cosmos, Universitat de Barcelona, Barcelona, Spain}
\orcid{0000-0002-8835-2910}
\author{Elies Gil-Fuster}
\affiliation{Dept. F\'{i}sica Qu\`{a}ntica i Astrof\'{i}sica, Universitat de 
Barcelona, Barcelona, Spain.}
\orcid{0000-0003-0411-9757}
\author{Jos\'{e} I. Latorre}
\affiliation{Barcelona Supercomputing Center}
\affiliation{Institut de Ci\`encies del Cosmos, Universitat de Barcelona, Barcelona, Spain}
\affiliation{Nikhef Theory Group, Science Park 105, 1098 XG Amsterdam, The Netherlands.}
\affiliation{Center for Quantum Technologies, National University of Singapore, Singapore.}
\orcid{0000-0003-1702-7018}

\begin{abstract}
A single qubit provides sufficient computational capabilities to construct a universal quantum classifier when assisted with a classical subroutine. This fact may be surprising since a single qubit only offers a simple superposition of two states and single-qubit gates only make a rotation in the Bloch sphere. The key ingredient to circumvent these limitations is to allow for multiple {\sl data re-uploading}. A quantum circuit can then be organized as a series of data re-uploading and single-qubit processing units. 
Furthermore, both data re-uploading and measurements can accommodate multiple dimensions in the input and several categories in the output, to conform to a universal quantum classifier. The extension of this idea to several qubits enhances the efficiency of the strategy as entanglement expands the superpositions carried along with the classification. Extensive benchmarking on different examples of the single- and multi-qubit quantum classifier validates its ability to describe and classify complex data.
\end{abstract}

\maketitle

\section{Introduction}

Quantum circuits that make use of a small number of quantum resources are of most importance to the field of quantum computation. Indeed, algorithms that need few qubits may prove relevant even if they do not attempt any quantum advantage, as they may be useful parts of larger circuits. 

A reasonable question to ask is what is the lower limit of quantum resources needed to achieve a given computation. A naive estimation for the quantum cost of a new proposed quantum algorithm is often made based on analogies with classical algorithms. But this may be misleading, as classical computation can play with memory in a rather different way as quantum computers do. The question then turns to the more refined problem of establishing the absolute minimum of quantum resources for a problem to be solved.

We shall here explore the power and minimal needs of quantum circuits assisted with a classical subroutine to carry out a general supervised classification task, that is, the minimum number of qubits, quantum operations and free parameters to be optimized classically. Three elements in the computation need renewed attention. The obvious first concern is to find a way to upload data in a quantum computer. Then, it is necessary to find the optimal processing of information, followed by an optimal measurement strategy. We shall revisit these three issues in turn. The non-trivial step we take here is to combine the first two, which is data uploading and processing.

There exist several strategies to design a quantum classifier. In general, they are inspired in well-known classical techniques such as artificial neural networks \cite{SSP14,WDKGK17,Juanjo} or kernel methods used in classical machine learning \cite{WBL12,RML14,ReviewQML,FN18,SBSW18,HCTHKCG19,SK19}. Some of these proposals \cite{WBL12,RML14,ReviewQML} encode the data values into a quantum state amplitude, which is manipulated afterward. These approaches need an efficient way to prepare and access to these amplitudes. State preparation algorithms are in general costly in terms of quantum gates and circuit depth, although some of these proposals use a specific state preparation circuit that only require few single-qubit gates. The access to the states that encode the data can be done efficiently by using a quantum random access memory (QRAM) \cite{QRAM}. However, this is experimentally challenging and the construction of a QRAM is still under development. Other proposals exploit hybrid quantum-classical strategies\cite{FN18,SBSW18,HCTHKCG19,SK19}. The classical parts can be used to construct the correct encoding circuit or as a minimization method to extract the optimal parameters of the quantum circuit, such as the angles of the rotational gates. In the first case, the quantum circuit computes the hardest instances of the classical classification algorithm as, for example, the inner products needed to obtain a kernel matrix. In the second case, the data is classified directly by using a \emph{parametrized quantum circuit}, whose variables are used to construct a cost function that should be minimized classically. This last strategy is more convenient for a Near Intermediate Scale Quantum computation (NISQ) since, in general, it requires short-depth circuits, and its variational core makes it more resistant to experimental errors. Our proposal belongs to this last category, the parametrized quantum classifiers.

A crucial part of a quantum classification algorithm is how data is encoded into the circuit. Proposals based on kernel methods design an encoding circuit which implements a feature map from the data space to the qubits Hilbert space. The construction of this quantum feature map may vary depending on the algorithm, but common strategies make use of the quantum Fourier transform or introduce data in multiple qubits using one- and two-qubit gates \cite{HCTHKCG19,SK19}. Both the properties of the tensor product and the entanglement generated in those encoding circuits capture the non-linearities of the data. In contrast, we argue that there is no need to use highly sophisticated encoding circuits nor a significant number of qubits to introduce these non-linearities. Single-qubit rotations applied multiple times along the circuit generate highly non-trivial functions of the data values. The main difference between our approach and the ones described above is that the circuit is not divided between the encoding and processing parts, but implements both multiple times along the algorithm.

Data re-uploading is considered as a manner of solving the limitations established by the \emph{no-cloning theorem}. Quantum computers cannot copy data, but classical devices can. For instance, a neural network takes the same input many times when processing the data in the hidden layer neurons. An analogous quantum neural network can only use quantum data once. Therefore, it makes sense to re-upload classical data along a quantum computation to bypass this limitation on the quantum circuit. By following this line of thought, we present an equivalence between data re-uploading and the Universal Approximation Theorem applied to artificial neural networks \cite{UAT}. Just as a network composed of a single hidden layer with enough neurons can reproduce any continuous function, a single-qubit classifier can, in principle, achieve the same by re-uploading the data enough times.

The single-qubit classifier illustrates the computational power that a single qubit can handle. This proposal is to be added to other few-qubit benchmarks in machine learning \cite{power1qubit}. The input redundancy has also been proposed to construct complex encoding in parametrized quantum circuits and in the construction of quantum feature maps \cite{inputredundancy, SK19}. These and other proposals mentioned in the previous paragraphs are focused on representing classically intractable or very complex kernel functions with few qubits. On the contrary,  the focus of this work is to distill the minimal amount of quantum resources, i.e., the number of qubits and gates, needed for a given classification task quantified in terms of the number of qubits and unitary operations. The main result of this work is, indeed, to show that there is a trade-off between the number of qubits needed to perform classification and multiple data re-uploading. That is, we may use fewer qubits at the price of re-entering data several times along the quantum computation.
 
We shall illustrate the power of a single- and multi-qubit classifiers with data re-uploading with a series of examples. First, we classify points in a plane that is divided into two areas. Then, we extend the number of regions on a plane to be classified. Next, we consider the classification of multi-dimensional patterns and, finally, we benchmark this quantum classifier with non-convex figures. For every example, we train a parametrized quantum circuit that carries out the task and we analyze its performance in terms of the circuit architecture, i.e., for single- and multi-qubit classifiers with and without entanglement between qubits.

This paper is structured as follows. First, in  Section \ref{sec:structure}, we present the basic structure of a single-qubit quantum classifier. Data and processing parameters are uploaded and re-uploaded using one-qubit general rotations. For each data point, the final state of the circuit is compared with the target state assigned to its class, and the free parameters of the circuit are updated accordingly using a classical minimization algorithm. Next, in Section \ref{sec:universal}, we motivate the data re-uploading approach by using the Universal Approximation Theorem of artificial neural networks. In Section \ref{sec:multiq}, we introduce the extension of this classifier to multiple qubits. Then, in Section \ref{sec:min}, we detail the minimization methods used to train the quantum classifiers. Finally, in Section \ref{sec:benchmark}, we benchmark single- and multi-qubit quantum classifiers defined previously with problems of different dimensions and complexity and compare their performance respect to classical classification techniques. The conclusions of this proposal for a quantum classifier are exposed in Section \ref{sec:conclusions}.

\section{Structure of a single-qubit quantum classifier\label{sec:structure}}

The global structure of any quantum circuit can be divided into three elements: uploading of information onto a quantum state, processing of the quantum state, and measurement of the final state. It is far from obvious how to implement each of these elements optimally to perform a specific operation. We shall now address them one at a time for the task of classification.

\subsection{Re-uploading classical information}

To load classical information onto a quantum circuit is a highly non-trivial task \cite{WBL12}. A critical example is the processing of big data. While there is no in-principle obstruction to upload large amounts of data onto a state, it is not obvious how to do it. 

The problem we address here is not related to a large amount of data. It is thus possible to consider a quantum circuit where all data are loaded in the coefficients of the initial wave function \cite{MNKF18,SBSW18,HCTHKCG19,inputredundancy,power1qubit}. In the simplest of cases, data are uploaded as rotations of qubits in the computational basis. A quantum circuit would then follow that should perform some classification.

This strategy would be insufficient to create a universal quantum classifier with a single qubit. A first limitation is that a single qubit only has two degrees of freedom, thus only allowing to represent data in a two-dimensional space. No quantum classifier in higher dimensions can be created if this architecture is to be used. A second limitation is that, once data is uploaded, the only quantum circuit available is a rotation in the  Bloch sphere. It is easy to prove that a single rotation cannot capture any non-trivial separation of patterns in the original data.

We need to turn to a different strategy, which turns out to be inspired by neural networks. In the case of feed-forward neural networks, data are entered in a network in such a way that they are processed by subsequent layers of neurons. The key idea is to observe that the original data are processed several times, one for each neuron in the first hidden layer. Strictly speaking, data are re-uploaded onto the neural network. If neural networks were affected by some sort of no-cloning theorem, they could not work as they do. Coming back to the quantum circuit, we need to design a new architecture where data can be introduced several times into the circuit.

\begin{figure}[t!]
\centering
\subfigure[\hspace{0.05cm} Neural network \label{Fig:nn}]{\includegraphics[width=0.18\textwidth]{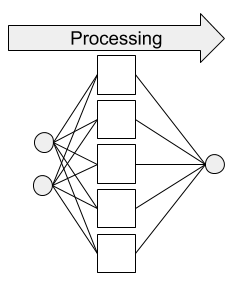}}
\subfigure[\hspace{0.05cm} Quantum classifier \label{Fig:q_sch}]{\includegraphics[width=0.28\textwidth]{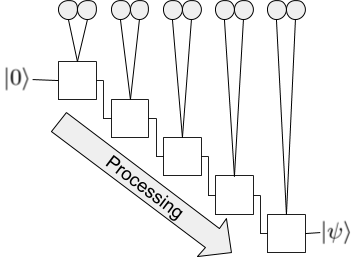}}
\caption{Simplified working schemes of a neural network and a single-qubit quantum classifier with data re-uploading. In the neural network, every neuron receives input from all neurons of the previous layer. In contrast with that, the single-qubit classifier receives information from the previous processing unit and the input (introduced classically). It processes everything all together and the final output of the computation is a quantum state encoding several repetitions of input uploads and processing parameters.}
\label{Fig:schemes}
\end{figure}

The central idea to build a universal quantum classifier with a single qubit is thus to re-upload classical data along with the computation. Following the comparison with an artificial neural network with a single hidden layer, we can represent this re-upload diagrammatically, as it is shown in Figure \ref{Fig:schemes}. Data points in a neural network are introduced in each processing unit, represented with squares, which are the neurons of the hidden layer. After the neurons process these data, a final neuron is necessary to construct the output to be analyzed. Similarly, in the single-qubit quantum classifier, data points are introduced in each processing unit, which this time corresponds to a unitary rotation. However, each processing unit is affected by the previous ones and re-introduces the input data. The final output is a quantum state to be analyzed as it will be explained in the next subsections.

The explicit form of this single-qubit classifier is shown in Figure \ref{Fig:VQC}. Classical data are re-introduced several times in a sequence interspaced with processing units. We shall consider the introduction of data as a rotation of the qubit. This means that data from three-dimensional space, $\vec x$, can be re-uploaded using unitaries that rotate the qubit $U(\vec{x})$. Later processing units will also be rotations as discussed later on. The whole structure needs to be trained in the classification of patterns. 

As we shall see, the performance of the single-qubit quantum classifier will depend on the number of re-uploads of classical data. This fact will be explored in the results section.

\subsection{Processing along re-uploading}

The single-qubit classifier belongs to the category of parametrized quantum circuits. The performance of the circuit is quantified by a figure of merit, some specific $\chi^2$ to be minimized and defined later. We need, though, to specify the processing gates present in the circuit in terms of a classical set of parameters.

\begin{figure}[t!]
\centering
\subfigure[\vspace{0.05cm}\hspace{0.05cm}Original scheme]{
\Qcircuit @C=0.85em @R=.9em  @!R{
 & & \mbox{$L(1)$} & & & & & & \mbox{$L(N)$} & &\\
\lstick{|0\rangle} & \gate{U\left(\vec{x}\right)} & \qw & \gate{U(\vec{\phi}_{1})} & \qw & \cdots & &  \gate{U\left(\vec{x}\right)} & \qw & \gate{U(\vec{\phi}_{N})} & \qw & \meter \\
\vspace{0cm}
\gategroup{2}{2}{2}{4}{.7em}{--}
\gategroup{2}{8}{2}{10}{.7em}{--}
}
}

\subfigure[\hspace{0.05cm}Compressed scheme]{
\Qcircuit @C=0.9em @R=.9em  @!R{
 & & & \mbox{$L(1)$} & & & & & & & \mbox{$L(N)$} & \\
\lstick{|0\rangle} & \qw & \qw & \gate{U\left(\vec{\phi}_{1},\vec{x}\right)} & \qw & \qw & \cdots & & \qw & \qw &  \gate{U\left(\vec{\phi}_{N},\vec{x}\right)} & \qw & \qw & \meter \\
\vspace{0cm}
\gategroup{2}{4}{2}{4}{.7em}{--}
\gategroup{2}{11}{2}{11}{.7em}{--}
}
}
\caption{Single-qubit classifier with data re-uploading. The quantum circuit is divided into layer gates $L(i)$, which constitutes the classifier building blocks. In the upper circuit, each of these layers is composed of a $U(\vec{x})$ gate, which uploads the data, and a parametrized unitary gate $U(\vec{\phi})$. We apply this building block $N$ times and finally compute a cost function that is related to the fidelity of the final state of the circuit with the corresponding target state of its class. This cost function may be minimized by tunning the $\vec{\phi}_{i}$ parameters. Eventually, data and tunable parameters can be introduced with a single unitary gate, as illustrated in the bottom circuit.}
\label{Fig:VQC}
\end{figure}
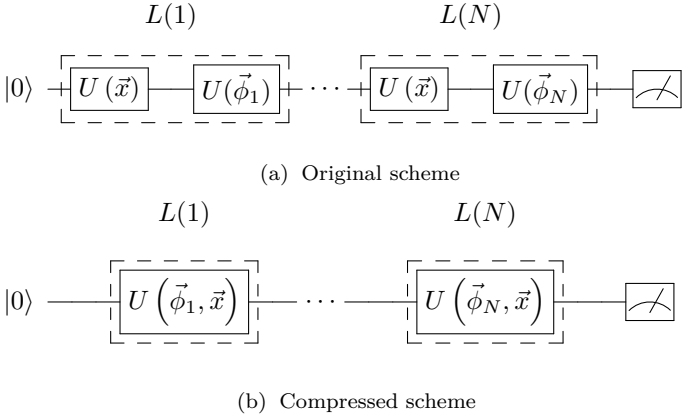

Given the simple structure of a single-qubit circuit presented in Figure \ref{Fig:VQC}, the data is introduced in a simple rotation of the qubit, which is easy to characterize. We just need to use arbitrary single-qubit rotations $U(\phi_{1},\phi_{2},\phi_{3})\in$ SU(2). We will write $U(\vec{\phi})$ with $\vec{\phi}=(\phi_{1},\phi_{2},\phi_{3})$. Then, the structure of the universal quantum classifier made with a single qubit is
\begin{equation}
\mathcal{U}(\vec{\phi}, \vec{x})\equiv U(\vec{\phi}_N) U(\vec x)\ldots U(\vec{\phi}_1) U(\vec x),
\end{equation}
which acts as
\begin{equation}
 |\psi \ra= \mathcal{U} (\vec{\phi}, \vec{x}) | 0\ra .
 \label{eq:qcircuit}
\end{equation}

The final classification of patterns will come from the results of measurements on $|\psi\ra$.
We may introduce the concept of {\sl processing layer} as the combination 
\begin{equation}
L(i)\equiv U({\vec \phi}_i) U(\vec x) ,
\end{equation}
so that the classifier corresponds to
\begin{equation}
\mathcal{U}(\vec{\phi},\vec{x})= L(N)\ldots L(1),
\label{eq:layers}
\end{equation}
where the depth of the circuit is $2N$. The more layers the more representation capabilities the circuit will have, and the more powerful the classifier will become. Again, this follows from the analogy to neural networks, where the size of the intermediate hidden layer of neurons is critical to represent complex functions. 

There is a way to compactify the quantum circuit into a shorter one. This can be done if we incorporate data and processing angles in a single step. Then, a layer would only need a single rotation to introduce data and tunable parameters, i.e. $L(i)=U(\vec{\phi},\vec{x})$. In addition, each data point can be uploaded with some weight $w_{i}$. These weights will play a similar role as weights in artificial neural networks, as we will see in the next section. Altogether, each layer gate can be taken as
\begin{equation}\label{eq:single_layer}
L(i)=U\left(\vec{\theta}_i+\vec{w}_{i}\circ\vec{x}\right),
\end{equation} 
where $\vec{w}_{i}\circ\vec{x}=\left(w_{i}^{1}x^{1},w_{i}^{2}x^{2},w_{i}^{3}x^{3}\right)$ is the Hadamard product of two vectors. In case the data points have dimension lesser than three, the rest of $\vec{x}$ components are set to zero. Such an approach reduces the depth of the circuit by half. Further combinations of layers into fewer rotations are also possible, but the nonlinearity inherent to subsequent rotations would be lost, and the circuit would not be performing well.

Notice that data points are introduced linearly into the rotational gate. Non-linearities will come from the structure of these gates. We chose this encoding function as we believe it is one of the lesser biased ways to encode data with unknown properties. Due to the structure of single-qubit unitary gates, we will see that this encoding is particularly suited for data with rotational symmetry. Still, it can also classify other kinds of data structures. We can also apply other encoding techniques, e.g. the ones proposed in Ref. \cite{SK19}, but for the scope of this work, we have just tested the linear encoding strategy as a proof of concept of the performance of this quantum classifier.

It is also possible to enlarge the dimensionality of the input space in the following way. Let us extend the definition of $i$-th layer to 
\begin{equation}
L(i)=U\left(\vec{\theta}_i^{(k)}+\vec{w}^{(k)}_i\circ\vec{x}^{(k)}\right)\cdots U\left(\vec{\theta}_i^{(1)}+\vec{w}^{(1)}_i\circ\vec{x}^{(1)}\right),
\label{eq:multiple_dim}
\end{equation}
where each data point is divided into $k$ vectors of dimension three. In general, each unitary $U$ could absorb as many variables as freedom in an SU(2) unitary. Each set of variables act at a time, and all of them have been shown to the circuit after $k$ iterations. Then, the layer structure follows. The complexity of the circuit only increases linearly with the size of the input space.

\subsection{Measurement}

The quantum circuit characterized by a series of processing angles $\{\theta_i\}$ and weights $\{w_{i}\}$ delivers a final state $|\psi\ra$, which needs to be measured. The results of each measurement are used to compute a $\chi^2$ that quantifies the error made in the classification. The minimization of this quantity in terms of the classical parameters of the circuit can be organized using any preferred supervised machine learning technique.

The critical point in the quantum measurement is to find an optimal way to associate outputs from the observations to target classes. The fundamental guiding principle to be used is given by the idea of maximal orthogonality of outputs \cite{Helstrom76}. This is easily established for a dichotomic classification, where one of two classes $A$ and $B$ have to be assigned to the final measurement of the single qubit. In such a case it is possible to measure the output probabilities $P(0)$ for $|0\ra$ and $P(1)$ for $|1\ra$. A given pattern could be classified into the $A$ class if $P(0)>P(1)$ and into $B$ otherwise. We may refine this criterium by introducing a bias. That is, the pattern is classified as $A$ if $P(0)>\lambda$, and as $B$ otherwise. The $\lambda$ is chosen to optimize the success of classification on a training set. Results are then checked on an independent validation set.

\begin{figure}
\includegraphics[width=.49\linewidth]{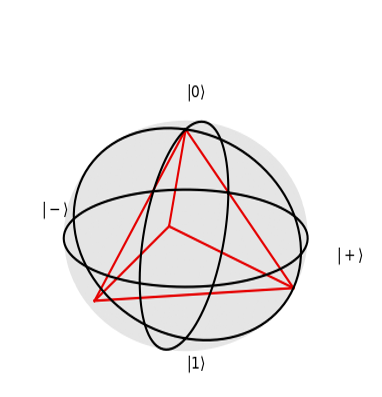}
\includegraphics[width=.49\linewidth]{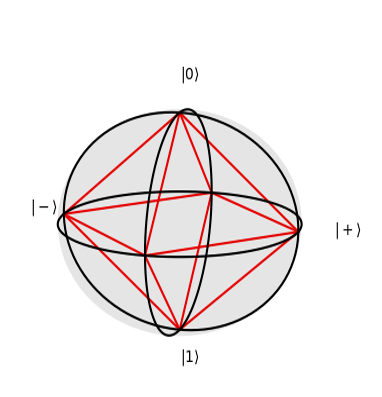}
\caption{Representation in the Bloch sphere of four and six maximally orthogonal points, corresponding to the vertices of a tetrahedron and an octahedron respectively. The single-qubit classifier will be trained to distribute the data points in one of these vertices, each one representing a class.}
\label{Fig:BlochSphere}
\end{figure}

The assignment of classes to the output reading of a single qubit becomes an involved issue when many classes are present. For the sake of simplicity, let us mention two examples for the case of classification to four distinct classes. One possible strategy consists on comparing the probability $P(0)$ to four sectors with three thresholds: $0\le \lambda_1 \le \lambda_2 \le \lambda_3 \le 1$. Then, the value of $P(0)$ will fall into one of them, and classification is issued. A second, more robust assignment is obtained by computing the overlap of the final state to one of the states of a label states-set. This states-set is to be chosen with maximal orthogonality among all of them. This second method needs from the maximally orthogonal points in the Bloch sphere. Figure \ref{Fig:BlochSphere} shows the particular cases that can be applied to a classification task of four and six classes. In general, a good measurement strategy may need some prior computational effort and refined tomography of the final state. Since we are proposing a single-qubit classifier, the tomography protocol will only require three measurements.

It is possible to interpret the single-qubit classifier in terms of geometry. The classifier opens a 2-dimensional Hilbert space, i.e., the Bloch sphere. As we encode data and classification within the parameters defining rotations, this Hilbert space is enough to achieve classification. Any operation $L(i)$ is a rotation on the Bloch sphere surface. With this point of view in mind, we can easily see that we can classify any point using only one unitary operation. We can transport any point to any other point on the Bloch sphere by nothing else than choosing the angles of rotation properly. However, this does not work for several data, as the optimal rotation for some data points could be very inconvenient for some other points. However, if more layers are applied, each one will perform a different rotation, and many different rotations together have the capability of enabling a feature map. Data embedded in this feature space can be easily separated into classes employing the regions on the Bloch sphere.

\subsubsection{A fidelity cost function}

We propose a very simple cost function motivated by the geometrical interpretation introduced above.  We want to force the quantum states $\ket{\psi(\vec\theta, \vec{w}, \vec{x})}$ to be as near as possible to one particular state on the Bloch sphere. The angular distance between the label state and the data state can be measured with the relative fidelity between the two states \cite{Nielsen11}. Thus, our aim is to maximize the average fidelity between the states at the end of the quantum circuit and the label states corresponding to their class. We define the following cost function that carries out this task,
\begin{equation}
\chi^2_{f}(\vec{\theta},\vec{w}) = \sum_{\mu=1}^{M} \left(1- |\la \tilde{\psi}_{s} | \psi(\vec{\theta}, \vec{w},\vec{x_\mu}) \ra |^2\right), 
\label{eq:fidelity_chi2}
\end{equation}
where $|\tilde{\psi}_{s}\ra$ is the correct label state of the $\mu$ data point, which will correspond to one of the classes.

\subsubsection{A weighted fidelity cost function}

We shall next define a refined version of the previous fidelity cost function to be minimized. The set of maximally orthogonal states in the Bloch sphere, i.e., the label states, are written as $\ket{\psi_c}$, where $c$ is the class. Each of these label states represents one class for the classifier. Now, we will follow the lead usually taken in neural network classification.

Let us define the quantity
\begin{equation}
F_{c}(\vec{\theta}, \vec{w}, \vec{x}) = |\bra{\tilde{\psi}_c}\psi(\vec{\theta}, \vec{w}, \vec{x})\ra|^2,
\label{eq:fidelity}
\end{equation}
where $M$ is the total number of training points, $\ket{\tilde{\psi}_c}$ is the label state of the class $c$ and $|\psi(\vec{\theta}, \vec{w}, \vec{x})\ra$ is the final state of the qubit at the end of the circuit. 
This fidelity is to be compared with the expected fidelity of a successful classification, $Y_{c}(\vec{x})$. For example, given a four-class classification and using the vertices of a tetrahedron as label states (as shown in Figure \ref{Fig:BlochSphere}), one expects $Y_{s}(\vec{x}) = 1$, where $s$ is the correct class, and $Y_{r}(\vec{x}) = 1/3$ for the other $r$ classes. In general, $Y_{c}(\vec{x})$ can be written as a vector with one entry equal to 1, the one corresponding to the correct class, and the others containing the overlap between the correct class label state and the other label states.

With these definitions, we can construct a cost function which turns out to be inspired by conventional cost functions in artificial neural networks. By weighting the fidelities of the final state of the circuit with all label states, we define the \textit{weighted fidelity} cost function as
\begin{equation}
\chi^2_{wf}(\vec{\alpha},\vec{\theta},\vec{w}) = \frac{1}{2} \sum_{\mu=1}^M \left(\sum_{c=1}^{\mathcal{C}}\left(\alpha_{c}F_{c}(\vec{\theta}, \vec{w}, \vec{x}_{\mu}) - Y_{c}(\vec{x}_{\mu})\right)^2\right),
\label{eq:conventional_chi2}
\end{equation}
where $M$ is the total number of training points, $\mathcal{C}$ is the total number of classes, $\vec{x}_{\mu}$ are the training points and $\vec{\alpha}=(\alpha_{1},\cdots,\alpha_{\mathcal{C}})$ are introduced as \emph{class weights} to be optimized together with $\vec{\theta}$ and $\vec{w}$ parameters. This weighted fidelity has more parameters than the fidelity cost function. These parameters are the weights for the fidelities. 

The main difference between the weighted fidelity cost function of Eq. \eqref{eq:conventional_chi2} and the fidelity cost function of Eq. \eqref{eq:fidelity_chi2} is how many overlaps do we need to compute. The $\chi_{wf}^2$ requires as many fidelities as classes every time we run the optimization subroutine, while the $\chi_{f}^2$ needs just one. This is not such a big difference for a few classes and only one qubit. It is possible to measure any state with a full tomography process which, for one qubit, is achievable. However, for many different classes, we expect that one measurement will be more efficient than many.

Besides the weighted fidelity cost function being costlier than the fidelity cost function, there is another qualitative difference between both. The fidelity cost function forces the parameters to reach the maximum in fidelities. Loosely speaking, this fidelity moves the qubit state to where it should be. The weighted fidelity forces the parameters to be close to a specified configuration of fidelities. It moves the qubit state to where it should be and moves it away from where it should not. Therefore, we expect that the weighted fidelity will work better than the fidelity cost function. Moreover, this extra cost in terms of the number of parameters of the weighted fidelity cost function will only affect the classical minimization part of the algorithm. In a sense, we are increasing the classical processing part to reduce the quantum resources required for the algorithm, i.e. the number of quantum operations (layers). This fact gain importance in the NISQ computation era.

\section{Universality of the single-qubit classifier \label{sec:universal}}

After analyzing several classification problems, we obtain evidence that the single-qubit classifier introduced above can approximate any classification function up to arbitrary precision. In this section, we provide the motivation for this statement based on the Universal Approximation Theorem (UAT) of artificial neural networks \cite{UAT}.

\subsection{Universal Approximation Theorem}

\emph{Theorem--} Let $I_{m}=[0,1]^m$ be the $m$-dimensional unit cube and $C(I_{m})$ the space of continuous functions in $I_{m}$. Let the function $\varphi:\mathbb{R}\rightarrow\mathbb{R}$ be a nonconstant, bounded and continuous function and $f:I_m\to\mathbb{R}$ a function. Then, for every $\epsilon>0$, there exists an integer $N$ and a function $h:I_m\to\mathbb{R}$, defined as
\begin{equation}
h(\vec{x})=\sum_{i=1}^N\alpha_{i} \ \varphi\left(\vec{w}_{i}\cdot\vec{x}+b_{i}\right),
\end{equation}
with $\alpha_{i},b_{i}\in\mathbb{R}$ and $\vec{w}_{i}\in\mathbb{R}^{m}$, such that $h$ is an approximate realization of $f$ with precision $\epsilon$, i.e.,
\begin{equation}
|h(\vec{x})-f(\vec{x})|<\epsilon
\end{equation}
for all $\vec{x}\in I_{m}$.

In artificial neural networks, $\varphi$ is the activation function, $\vec{w}_{i}$ are the weights for each neuron, $b_{i}$ are the biases and $\alpha_{i}$ are the neuron weights that construct the output function. Thus, this theorem establishes that it is possible to reconstruct any continuous function with a single layer neural network of $N$ neurons. The proof of this theorem for the sigmoidal activation function can be found in Ref. \cite{Cybenko89}. This theorem was generalized for any nonconstant, bounded and continuous activation function in Ref. \cite{UAT}. Moreover, Ref. \cite{UAT} presents the following corollary of this theorem: $\varphi$ could be a nonconstant finite linear combination of periodic functions, in particular, $\varphi$ could be a nonconstant trigonometric polynomial.

\subsection{Universal Quantum Circuit Approximation}

The single-qubit classifier is divided into several layers which are general SU(2) rotational matrices. There exist many possible decompositions of an SU(2) rotational matrix. In particular, we use
\begin{equation}
U(\vec{\phi})=U(\phi_{1},\phi_{2},\phi_{3})= e^{i\phi_2\sigma_{z}}e^{i\phi_{1}\sigma_{y}}e^{i\phi_{3}\sigma_{z}},
\end{equation} 
where $\sigma_i$ are the conventional Pauli matrices.
Using the SU(2) group composition law, we can rewrite the above parametrization in a single exponential,
\begin{equation}
U(\vec{\phi})=e^{i \vec{\omega}(\vec{\phi})\cdot\vec{\sigma}},
\end{equation}
with $\vec{\omega}(\vec{\phi})=\left(\omega_{1}(\vec{\phi}),\omega_{2}(\vec{\phi}),\omega_{3}(\vec{\phi})\right)$ and
\begin{align}
\omega_{1}(\vec{\phi})&= d \  \mathcal{N}\sin\left((\phi_{2}-\phi_{3})/2\right)\sin\left(\phi_{1}/2\right),\\
\omega_{2}(\vec{\phi})&= d \  \mathcal{N}\cos\left((\phi_{2}-\phi_{3})/2\right)\sin\left(\phi_{1}/2\right) , \\ 
\omega_{3}(\vec{\phi})&= d \ \mathcal{N}\sin\left((\phi_{2}+\phi_{3})/2\right)\cos\left(\phi_{1}/2\right),
\end{align}
where $\mathcal{N}=\left(\sqrt{1-\cos^2d} \right)^{-1}$ and $\cos d=\cos\left((\phi_{2}+\phi_{3})/2\right)\cos\left(\phi_{1}/2\right)$.

The single-qubit classifier codifies the data points into $\vec{\phi}$ parameters of the $U$ unitary gate. In particular, we can re-upload data together with the tunable parameters as defined in Eq. \eqref{eq:single_layer}, i.e. 
\begin{equation}
\vec{\phi}(\vec{x}) =(\phi_{1}(\vec{x}),\phi_{2}(\vec{x}),\phi_{3}(\vec{x}))= \vec{\theta} + \vec{w}\circ\vec{x}.
\end{equation} 
Thus,
\begin{equation}
\mathcal{U}(\vec{x})=U_{N}(\vec{x})U_{N-1}(\vec{x})\cdots U_{1}(\vec{x})=\prod_{i=1}^{N}e^{i\vec{\omega}(\vec{\phi}_{i}(\vec{x}))\cdot\vec{\sigma}},
\end{equation}
Next, we apply the Baker-Campbell-Hausdorff (BCH) formula \cite{BCH} to the above equation,
\begin{equation}
\mathcal{U}(\vec{x}) = \exp\left[i\sum_{i=1}^{N}\vec{\omega}(\vec{\phi}_{i}(\vec{x}))\cdot\vec{\sigma} + \mathcal{O}_{corr}\right].
\end{equation}
Notice that the remaining BCH terms $\mathcal{O}_{corr}$ are also proportional to Pauli matrices due to $[\sigma_{i},\sigma_{j}]=2i\epsilon_{ijk}\sigma_{k}$.

Each $\vec{\omega}$ terms are trigonometric functions, unconstant, bounded and continuous. Then
\begin{multline}
\sum_{i=1}^{N}\vec{\omega}(\vec{\phi}_{i}(\vec{x}))=\sum_{i=1}^{N}\left(\omega_1(\vec{\phi}_{i}(\vec{x})),\omega_2(\vec{\phi}_{i}(\vec{x})),\omega_3(\vec{\phi}_{i}(\vec{x}))\right) \\
 = \sum_{i=1}^{N}\left(\omega_{1}(\vec{\theta}_{i} + \vec{w}_{i}\circ\vec{x}),\omega_{2}(\vec{\theta}_{i} + \vec{w}_{i}\circ\vec{x}),\omega_{3}(\vec{\theta}_{i} + \vec{w}_{i}\circ\vec{x})\right) \\
 = \left(f_{1}(\vec{x}),f_2(\vec{x}),f_3(\vec{x})\right).
\end{multline}

We still have to deal with the remaining terms $\mathcal{O}_{corr}$ of the BCH expansion. Instead of applying such expansion, we can use again the SU(2) group composition law to obtain the analytical formula of $\mathcal{U}(\vec{x})= e^{i\vec{\xi}\left(\vec{x}\right)\cdot\vec{\sigma}}$, where $\vec{\xi}(\vec{x})$ will be an inextricably trigonometric function of $\vec{x}$. The $\mathcal{O}_{corr}$ terms are proportional to $\vec{\sigma}$ matrices, so $\mathcal{O}_{corr} = \vec{\varrho}(\vec{x})\cdot\vec{\sigma}$ for some function $\vec{\varrho}(\vec{x})$. Then,
\begin{equation}
\mathcal{U}(\vec{x}) = e^{i\vec{\xi}(\vec{x})\cdot\vec{\sigma}} = e^{i\vec{f}(\vec{x})\cdot\vec{\sigma} + i\vec{\varrho}(\vec{x})\cdot\vec{\sigma}}.
\end{equation}
Thus, $\mathcal{O}_{corr}$ terms can be absorbed in $\vec{f}(\vec{x})$.

For each data point $\vec{x}$, we obtain a final state that will contain these $\vec{\xi}(\vec{x})$ functions. With all training points, we construct a cost function that can include new parameters $\alpha_{c}$ for each class if we use the weighted fidelity cost function of Eq. \eqref{eq:conventional_chi2}. The function obtained from the combination of $\vec{\xi}(x)$ and $\alpha_{c}$ is expected to be complex enough to probably represent almost any continuous function. However, more parameters are necessary to map this argument with the UAT expression.

If we compare the parameters of the UAT with the single-qubit circuit parameters, the $\vec{w}_{i}$ will correspond with the weights, the $\vec{\theta_{i}}$ with the biases $b_{i}$, the number of layers $N$ of the quantum classifier will correspond with the number of neurons in the hidden layer and $\vec{\omega}$ functions with the activation functions $\varphi$. 

We have explained why it is necessary to re-upload the data at each layer and why a single qubit could be a universal classifier. As has been stated before, an artificial neural network introduces the data points in each hidden neuron, weights them and adds some bias. Here we cannot just copy each data point because the non-cloning theorem, so we have to re-upload it at each layer. 

\section{From single- to multi-qubit quantum classifier \label{sec:multiq}}

The single-qubit classifier cannot carry any quantum advantage respect classical classification techniques such as artificial neural networks. In the previous sections, we have defined a quantum mechanical version of a neural network with a single hidden layer. In general, a huge amount of hidden neurons is necessary to approximate a target function with a single layer. To circumvent this inconvenience, more hidden layers are introduced, leading eventually to the concept of \emph{deep neural networks}.

By using the single-qubit classifier formalism that we have introduced in the previous sections, we propose its generalization to more qubits. The introduction of multiple qubits to this quantum classifier may improve its performance as more hidden layers improve the classification task of an artificial neural network. With the introduction of entanglement between these qubits, we reduce the number of layers of our classifier as well as propose a quantum classification method that can achieve quantum advantage.

Figure \ref{Fig:schemes} shows the analogy between a neural network with a single hidden layer and a single-qubit classifier. The generalization of this analogy is not so obvious. A multi-qubit classifier without entanglement could have some similarities with a convolutional neural network, where each qubit could represent a neural network by itself. However, it is not clear if the introduction of entanglement between qubits can be understood as a deep neural network architecture. The discussion around this analogy as well as an extended study of the performance of a multi-qubit classifier is beyond the scope of this work. In the next subsections, we present a general proposal for a multi-qubit classifier which we compare with the single-qubit one in Section \ref{sec:benchmark}.

\subsection{Measurement strategy and cost function for a multi-qubit classifier}

With a single-qubit classifier, the measurement strategy consisting on comparing the final state of the circuit with a pre-defined target state was achievable. Experimentally, one needs to perform a quantum state tomography protocol of only three measurements. However, if more qubits are to be considered, tomography protocols become exponentially expensive in terms of number of measurements. 

We propose two measurement strategies for a multi-qubit classifier. The first one is the natural generalization of the single-qubit strategy, although it will become unrealizable for a large number of qubits. We compare the final state of the circuit with one of the states of the computational basis, one for each class. The second strategy consist on focusing in one qubit and depending on its state associate one or other class. This is similar to previous proposals of binary multi-qubit classifiers \cite{FN18}, although we add the possibility of multiclass classification by introducing several thresholds (see Section \ref{sec:structure}).

Another part that should be adapted is the definition of the cost function. In particular, we use different functions for each strategy explained above. 

For the first strategy, we use the fidelity cost function of Eq. \eqref{eq:fidelity_chi2}. Its generalization to more qubits is straightforward. However, the orthogonal states used for a multi-qubit classifier are taken as the computational basis states. A more sophisticated set of states could be considered to improve the performance of this method.

For the second strategy, we use the weighted fidelity cost function. As stated above, we just focus on one qubit, thus
\begin{equation}
F_{c,q}(\vec{\theta}, \vec{w}, \vec{x}) = \bra{\tilde{\psi}_c}\rho_{q}(\vec{\theta}, \vec{w}, \vec{x})\ket{\tilde{\psi}_c},
\label{eq:fidelity_multiq}
\end{equation}
where $\rho_{q}$ is the reduced density matrix of the qubit to be measured.
Then, the weighted fidelity cost function can be adapted as
\begin{multline}
\chi_{wf}^2(\vec{\alpha},\vec{\theta},\vec{w}) = \\ \frac{1}{2} \sum_{\mu=1}^M \sum_{c=1}^{\mathcal{C}}\left(\sum_{q=1}^{Q}\left(\alpha_{c,q}F_{c,q}(\vec{\theta}, \vec{w}, \vec{x}_{\mu}) - Y_{c}(\vec{x}_{\mu})\right)^2\right),
\label{eq:conventional_chi2_multiq}
\end{multline}
where we average over all $Q$ qubits that form the classifier. Eventually, we can just measure one of these qubits, reducing the number of parameters to be optimized.

\subsection{Quantum circuits examples}

\begin{figure}[t!]
\centering
\subfigure[\hspace{0.05cm} Ansatz with no entanglement]{
\Qcircuit @C=0.85em @R=.9em  @!R{
\lstick{|0\rangle} & \qw & \gate{L_{1}(1)} & \gate{L_{1}(2)} & \gate{L_{1}(3)} & \qw & \cdots & & \gate{L_{1}(N)} & \qw \\
\lstick{|0\rangle} & \qw & \gate{L_{2}(1)} & \gate{L_{2}(2)} & \gate{L_{2}(3)} & \qw & \cdots & & \gate{L_{2}(N)} & \qw \\
\vspace{0cm}
}
}

\subfigure[\hspace{0.05cm} Ansatz with entanglement]{
\Qcircuit @C=0.85em @R=.9em  @!R{
\lstick{|0\rangle} & \qw & \gate{L_{1}(1)} & \ctrl{1} & \gate{L_{1}(2)} & \ctrl{1} & \qw & \cdots & & \ctrl{1} & \gate{L_{1}(N)} & \qw  \\
\lstick{|0\rangle} & \qw & \gate{L_{2}(1)} & \ctrl{-1} & \gate{L_{2}(2)} & \ctrl{-1} & \qw & \cdots & & \ctrl{-1} & \gate{L_{2}(N)} & \qw \\
\vspace{0cm}
\gategroup{1}{3}{2}{4}{.7em}{--}
\gategroup{1}{5}{2}{6}{.7em}{--}
\gategroup{1}{11}{2}{11}{.7em}{--}
}
}
\caption{Two-qubit quantum classifier circuit without entanglement (top circuit) and with entanglement (bottom circuit). Here, each layer includes a rotation with data re-uploading in both qubits plus a CZ gate if there is entanglement. The exception is the last layer, which does not have any CZ gate associated to it. For a fixed number of layers, the number of parameters to be optimized doubles the one needed for a single-qubit classifier.}
\label{Fig:2qubit_circuit}
\end{figure}
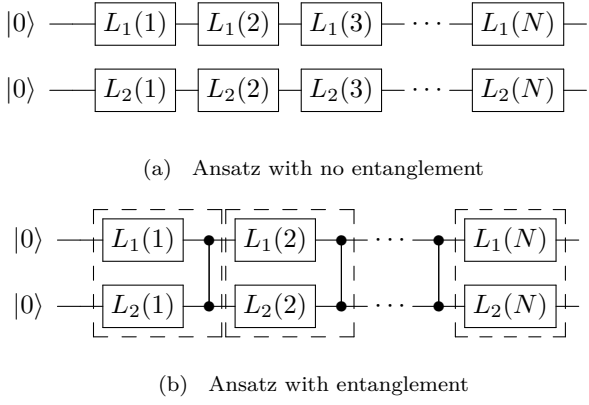

The definition of a multi-qubit quantum classifier circuit could be as free as is the definition of a multi-layer neural network. In artificial neural networks, it is far from obvious what should be the number of hidden layers and neurons per layer to perform some task. Besides, it is, in general, problem-dependent. For a multi-qubit quantum classifier, there is extra degree of freedom in the circuit-design: how to introduce the entanglement. This is precisely an open problem in parametrized quantum circuits: to find a correct ansatz for the entangling structure of the circuit.

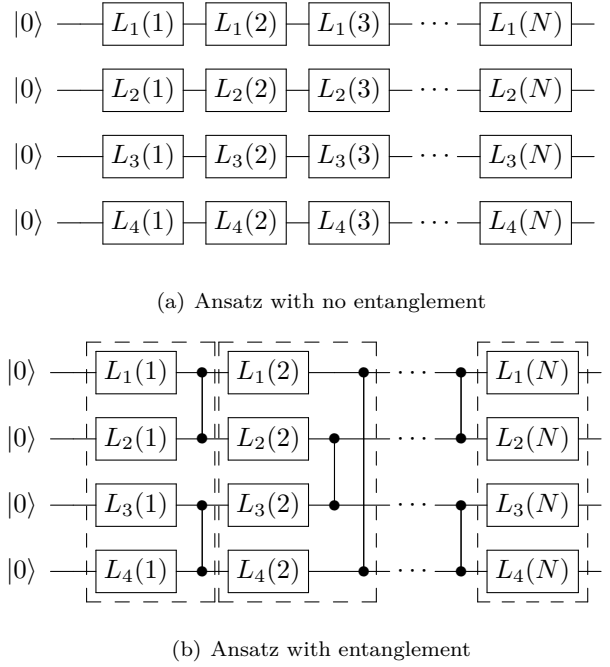
\begin{figure}[t!]
\centering
\subfigure[Ansatz with no entanglement]{
\Qcircuit @C=0.85em @R=.9em  @!R{
\lstick{|0\rangle} & \qw & \gate{L_{1}(1)} & \gate{L_{1}(2)} & \gate{L_{1}(3)} & \qw & \cdots & & \gate{L_{1}(N)} & \qw \\
\lstick{|0\rangle} & \qw & \gate{L_{2}(1)} & \gate{L_{2}(2)} & \gate{L_{2}(3)} & \qw & \cdots & & \gate{L_{2}(N)} & \qw \\
\lstick{|0\rangle} & \qw & \gate{L_{3}(1)} & \gate{L_{3}(2)} & \gate{L_{3}(3)} & \qw & \cdots & & \gate{L_{3}(N)} & \qw \\
\lstick{|0\rangle} & \qw & \gate{L_{4}(1)} & \gate{L_{4}(2)} & \gate{L_{4}(3)} & \qw & \cdots & & \gate{L_{4}(N)} & \qw \\
\vspace{0cm}
}
}

\subfigure[Ansatz with entanglement]{
\Qcircuit @C=0.85em @R=.9em  @!R{
\lstick{|0\rangle} & \qw & \gate{L_{1}(1)} & \ctrl{1} & \gate{L_{1}(2)} & \qw & \ctrl{3} & \qw & \cdots & & \ctrl{1} & \gate{L_{1}(N)} & \qw  \\
\lstick{|0\rangle} & \qw & \gate{L_{2}(1)} & \ctrl{-1} & \gate{L_{2}(2)} & \ctrl{1} & \qw & \qw & \cdots & & \ctrl{-1} & \gate{L_{2}(N)} & \qw \\
\lstick{|0\rangle} & \qw & \gate{L_{3}(1)} & \ctrl{1} & \gate{L_{3}(2)} & \ctrl{-1} & \qw &  \qw & \cdots & & \ctrl{1} & \gate{L_{3}(N)} & \qw \\
\lstick{|0\rangle} & \qw & \gate{L_{4}(1)} & \ctrl{-1} & \gate{L_{4}(2)} & \qw & \ctrl{-3} & \qw & \cdots & & \ctrl{-1} & \gate{L_{4}(N)} & \qw \\
\vspace{0cm}
\gategroup{1}{3}{4}{4}{.7em}{--}
\gategroup{1}{5}{4}{7}{.7em}{--}
\gategroup{1}{12}{4}{12}{.7em}{--}
}
}
\caption{Four-qubit quantum classifier circuits. Without entanglement (top circuit), each layer is composed by four parallel rotations. With entanglement (bottom circuit) each layer includes a parallel rotation and two parallel CZ gates. The order of CZ gates alternates in each layer between (1)-(2) and (3)-(4) qubits and (2)-(3) and (1)-(4) qubits. The exception is in the last layer, which does not contain any CZ gate. For a fixed number of layers, the number of parameters to be optimized quadruples the ones needed for a single-qubit classifier.}
\label{Fig:4qubit_circuit}
\end{figure}

Figures \ref{Fig:2qubit_circuit} and \ref{Fig:4qubit_circuit} show the explicit circuits used in this work. For a two-qubit classifier without entanglement, and similarly for a four-qubit classifier, we identify each layer as parallel rotations on all qubits. We introduce the entanglement using CZ gates between rotations that are absorbed in the definition of layer. For two-qubit classifier with entanglement, we apply a CZ gate after each rotation with exception of the last layer. For a four-qubit classifier, two CZ gates are applied after each rotation alternatively between (1)-(2) and (3)-(4) qubits and (2)-(3) and (1)-(4) qubits.

The number of parameters needed to perform the optimization doubles the ones needed for a single-qubit classifier for the two-qubit classifier and quadruples for the four-qubit classifier. For $N$ layers, the circuit depth is $N$ for the non-entangling classifiers and $2N$ for the entangling classifiers.

\section{Minimization methods \label{sec:min}}

The practical training of a parametrized single-qubit or multi-qubit quantum classifier needs minimization in the parameter space describing the circuit. This is often referred as a hybrid algorithm, where classical and quantum logic coexist and benefit from one another. To be precise, the set of $\{\theta_i\}$ angles and $\{w_{i}\}$ weights, together with $\alpha_{q,l}$ parameters if applicable, forms a space to be explored in search of a minimum $\chi^2$. In parameter landscapes as big as the ones treated here, or in regular neural network classification, the appearance of local minima is ultimately unavoidable. The composition of rotation gates renders a large product of independent trigonometric functions. It is thus clear to see that our problem will be overly populated with minima. The classical minimizer can easily get trapped in a not optimal one.

Our problem is reduced to minimizing a function of many parameters. For a single-qubit classifier, the number of parameters is $(3+d)N$ where $d$ is the dimension of the problem, i.e. the dimension of $\vec{x}$, and $N$ is the number of layers. Three of these parameters are the rotational angles and the other $d$ correspond with the $\vec{w}_{i}$ weight. If using the weighted fidelity cost function, we should add $\mathcal{C}$ extra parameters, one for each class. 

In principle, one does not know how is the parameter landscape of the cost function to be minimized. If the cost function were, for example, a convex function, a downhill strategy would be likely to work properly. The pure downhill strategy is known as gradient descent. In machine learning, the method commonly used is a Stochastic Gradient Descent (SGD) \cite{Nielsen15}.

There is another special method of minimization known as L-BFGS-B \cite{BLNZ95}. This method has been used in classical machine learning with very good results \cite{scikit-learn}.

The results we present from now on are starred by the L-BFGS-B algorithm, as we found it is accurate and relatively fast. We used open source software \cite{scipy} as the core of the minimization with own made functions to minimize. The minimizer is taken as a black box whose parameters are set by default. As this is the first attempt of constructing a single- or multi-qubit classifier, further improvements can be done on the hyperparameters of minimization. 

Nevertheless we have also tested a SGD algorithm for the fidelity cost function. This whole algorithm has been developed by us following the steps from \cite{Nielsen11}. The details can be read in Appendix \ref{sec:SGD}. In general, we found that L-BFGS-B algorithm is better than SGD. This is something already observed in classical neural networks. When the training set is small, it is often more convenient to use a L-BFGS-B strategy than a SGD. We were forced to use small training sets due to computational capabilities for our simulations. Numerical evidences on this arise when solving the problems we face for these single- and multi-qubit classifiers with classical standard machine learning libraries \cite{scikit-learn}.
This can be understood with a simple argument. Neural networks or our quantum classifier are supposed to have plenty of local minima. Neural networks have got huge products of non linear functions. The odds of having local minima are then large. In the quantum circuits side, there are nothing but trigonometric functions. In both cases, if there are a lot of training points it is more likely to find some of them capable of getting us out of local minima. If this is the case, SGD is more useful for being faster. On the contrary, when the training set is small, we have to pick an algorithm less sensitive to local minima, such as the L-BFGS-B.

\section{Benchmark of a single- and multi-qubit classifier \label{sec:benchmark}}

We can now tackle some classification problems. We will prove that a single-qubit classifier can perform a multi-class classification for multi-dimensional data and that a multi-qubit classifier, in general, improves these results.

We construct several classifiers with different number of layers. We then train the circuits with a training set of random data points to obtain the values of the free parameters $\{\theta_{i}\}$ and $\{w_{i}\}$ for each layer and $\{\alpha_{i}\}$ when applicable. We use the cost functions defined in Eq. \eqref{eq:conventional_chi2} and Eq. \eqref{eq:fidelity_chi2}. Then, we test the performance of each classifier with a test set independently generated and one order of magnitud greater than the training set. For the sake of reproducibility, we have fixed the same seed to generate all data points. For this reason, the test and training set points are the same for all problems. For more details, we provide the explicit code used in this work \cite{codigo}.

We run a single-, two- and four-qubit classifiers, with and without entanglement, using the two cost functions described above. We benchmark several classifiers formed by $L=1,2,3,4,5,6,8$ and 10 layers. 

In the following subsections, we describe the particular problems addressed with these single- and multi-qubit classifiers with data re-uploading. We choose four problem types: a simple binary classification, a classification of a figure with multiple patterns, a multi-dimensional classification and a non-convex figure.

The code used to define and benchmark the single- and multi-qubit quantum classifier is open and can be found in Ref. \cite{codigo}.

\subsection{Simple example: classification of a circle}

\begin{table*}[t!]
\centering
\begin{tabular}{c|c|cc|c|cc|cc}
  & \multicolumn{3}{c|}{$\chi^{2}_{f}$} & \multicolumn{5}{c}{$\chi^{2}_{wf}$} \\
 \hline
Qubits & 1 & \multicolumn{2}{c|}{2 } & 1 & \multicolumn{2}{c|}{2} & \multicolumn{2}{c}{4 }   \\
Layers  & & No Ent. & Ent. & & No Ent. & Ent. & No Ent. & Ent. \\ 
 \hline
 1 & 0.50 & 0.75 & -- & 0.50 & 0.76 & -- & 0.76 & --  \\
 2 & 0.85 & 0.80 & 0.73 & 0.94 & 0.96 & 0.96 & 0.96 & 0.96  \\
 3 & 0.85 & 0.81 & 0.93 & 0.94 & 0.97 & 0.95 & 0.97 & 0.96  \\
 4 & 0.90 & 0.87 & 0.87 & 0.94 & 0.97 & 0.96 & 0.97 & 0.96  \\
 5 & 0.89 & 0.90 & 0.93 & 0.96 & 0.96 & 0.96 & 0.96 & 0.96  \\
 6 & 0.92 & 0.92 & 0.90 & 0.95 & 0.96 & 0.96 & 0.96 & 0.96  \\
 8 & 0.93 & 0.93 & 0.96 & 0.97 & 0.95 & 0.97 & 0.95 & 0.96  \\
10 & 0.95 & 0.94 & 0.96 & 0.96 & 0.96 & 0.96 & 0.96 & 0.97  \\
\end{tabular}
\caption{Results of the single- and multi-qubit classifiers with data re-uploading for the circle problem. Numbers indicate the success rate, i.e. number of data points classified correctly over total number of points. Words ``Ent." and ``No Ent." refer to considering entanglement between qubits or not, respectively. We have used the L-BFGS-B minimization method with the weighted fidelity and fidelity cost functions. For this problem, both cost functions lead to high success rates. The multi-qubit classifier increases this success rate but the introduction of entanglement does not affect it significantly.}
\label{tab:results_circle}
\end{table*}

Let us start with a simple example. We create a random set of data on a plane with coordinates $\vec{x}=(x_{1},x_{2})$ with $x_{i}\in[-1,1]$. Our goal is to classify these points according to $x_{1}^2+x_{2}^2< r^2$, i.e. if they are inside or outside of a circle of radius $r$. The value of the radius is chosen in such a way that the areas inside and outside it are equal, that is, $r= \sqrt{\frac{2}{\pi}}$, so the probability of success if we label each data point randomly is 50\%. We create a train dataset with 200 random entries. We then validate the single-qubit classifier against a test dataset with 4000 random points.
 
The results of this classification are written in Table \ref{tab:results_circle}. With the weighted fidelity cost function, the single-qubit classifier achieves more than 90\% of success with only two layers, that is, 12 parameters. The results are worse with the fidelity cost function. For a two-qubit and a four-qubit classifier, two layers are required to achieve 96\% of success rate, that is, 22 parameters for the two-qubit and 42 for the four-qubit. The introduction of entanglement does not change the result in any case. The results show a saturation of the success rate. Considering more layers or more qubits does not change this success rate.

The characterization of a closed curved is a hard problem for an artificial neural network that works in a linear regime, although enough neurons, i.e. linear terms, can achieve a good approximation to any function. On the contrary, the layers of a single-qubit classifier are rotational gates, which have an intrinsic non-linear behavior. In a sense, a circle becomes an easy function to classify as a linear function is for an artificial neural network. The circle classification is, in a sense, trivial for a quantum classifier. We need to run these classifiers with more complex figures or problems to test their performance.

\begin{figure}[t!]
\centering
\subfigure[\hspace{0.05cm} 1 layer]{\includegraphics[width=0.22\textwidth]{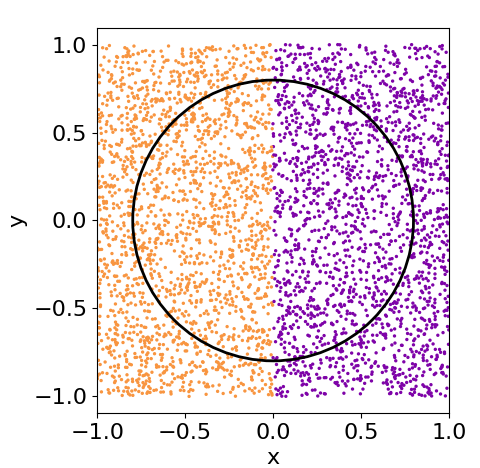}}
\subfigure[\hspace{0.05cm} 2 layers]{\includegraphics[width=0.22\textwidth]{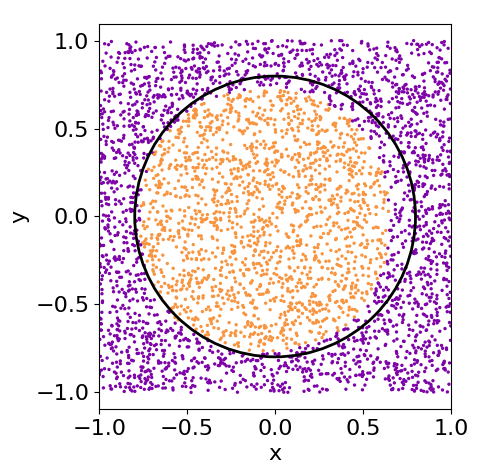}}
\subfigure[\hspace{0.05cm} 4 layers]{\includegraphics[width=0.22\textwidth]{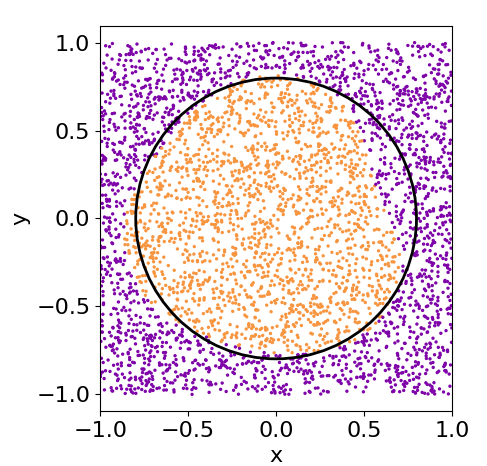}}
\subfigure[\hspace{0.05cm} 8 layers]{\includegraphics[width=0.22\textwidth]{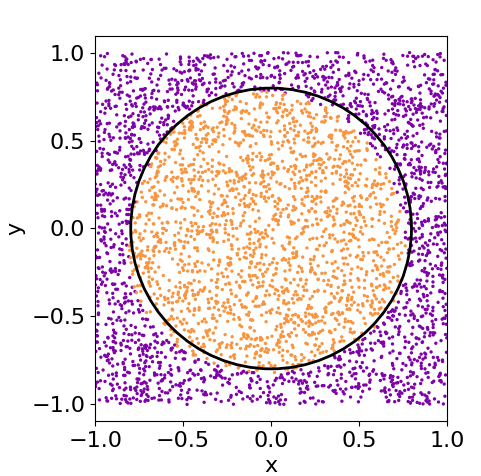}}
\caption{Results of the circle classification obtained with a single-qubit classifier with different number of layers using the L-BFGS-B minimizer and the weighted fidelity cost function. With one layer, the best that the classifier can do is to divide the plane in half. With two layers, it catches the circular shape which is readjusted as we consider more layers.}
\label{Fig:1circle_evol}
\end{figure}

It is interesting to compare classifiers with different number of layers. Figure \ref{Fig:1circle_evol} shows the result of the classification for a single-qubit classifier of 1, 2, 4 and 8 layers. As with only one layer the best classification that can be achieved consist on dividing the plane in half, with two layers the classifier catches the circular shape. As we consider more layers, the single-qubit classifier readjust the circle to match the correct radius.

\begin{table*}[t!]
\centering
\begin{tabular}{c|c|cc|c|cc|cc}
  & \multicolumn{3}{c|}{$\chi^{2}_{f}$} & \multicolumn{5}{c}{$\chi^{2}_{wf}$} \\
 \hline
Qubits & 1 & \multicolumn{2}{c|}{2 } & 1 & \multicolumn{2}{c|}{2} & \multicolumn{2}{c}{4 }   \\
Layers  & & No Ent. & Ent. & & No Ent. & Ent. & No Ent. & Ent. \\ 
 \hline
 1 & 0.73 & 0.56 & -- & 0.75 & 0.81 & -- & 0.88 & -- \\
 2 & 0.79 & 0.77 & 0.78 & 0.76 & 0.90 & 0.83 & 0.90 & 0.89 \\
 3 & 0.79 & 0.76 & 0.75 & 0.78 & 0.88 & 0.89 & 0.90 & 0.89 \\
 4 & 0.84 & 0.80 & 0.80 & 0.86 & 0.84 & 0.91 & 0.90 & 0.90 \\
 5 & 0.87 & 0.84 & 0.81 & 0.88 & 0.87 & 0.89 & 0.88 & 0.92 \\
 6 & 0.90 & 0.88 & 0.86 & 0.85 & 0.88 & 0.89 & 0.89 & 0.90 \\
 8 & 0.89 & 0.85 & 0.89 & 0.89 & 0.91 & 0.90 & 0.88 & 0.91 \\
10 & 0.91 & 0.86 & 0.90 & 0.92 & 0.90 & 0.91 & 0.87 & 0.91 \\
\end{tabular}
\caption{Results of the single- and multi-qubit classifiers with data re-uploading for the 3-circles problem. Numbers indicate the success rate, i.e. number of data points classified correctly over total number of points. Words ``Ent." and ``No Ent." refer to considering entanglement between qubits or not, respectively. We have used the L-BFGS-B minimization method with the weighted fidelity and fidelity cost functions. Weighted fidelity cost function presents better results than the fidelity cost function. The multi-qubit classifier reaches 0.90 success rate with a lower number of layers than the single-qubit classifier. The introduction of entanglement slightly increases the success rate respect the non-entangled circuit.}
\label{tab:results_3circles}
\end{table*}

\subsection{Classification of multiple patterns}

We want to show now that the single-qubit classifier can solve multi-class problems. We divide a 2D plane into several regions and assign a label to each one. We propose the following division: three regions corresponding to three circular sectors and the intermediate space between them. We call this problem the \textit{3-circles} problem. This is a hardly non-linear problem and, consequently, difficult to solve for a classical neural network in terms of computational power.

Table \ref{tab:results_3circles} shows the results for this four-class problem. For a single-qubit classifier, a maximum of 92\% of success is achieved with 10 layers, i.e. 54 parameters. From these results, it seems that this problem also saturates around 91\% of success. However, the introduction of more qubits and entanglement makes possible this result possible with less parameters. For two qubits with entanglement, 4 layers are necessary to achieve the same success as with a single-qubit, i.e. 34 parameters. For four qubits without entanglement 4 layers are also required. Notice also that, although the number of parameters increases significantly with the number of qubits, some of the effective operations are performed in parallel.

There is an effect that arises from this more complex classification problem: local minima. Notice that the success rate can decrease when we add more layers into our quantum classifier.  

\begin{figure}[t!]
\centering
\subfigure[\hspace{0.05cm} 1 layer]{\includegraphics[width=0.22\textwidth]{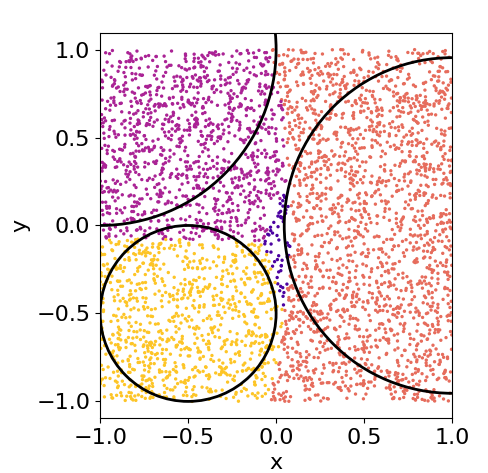}}
\subfigure[\hspace{0.05cm} 3 layers]{\includegraphics[width=0.22\textwidth]{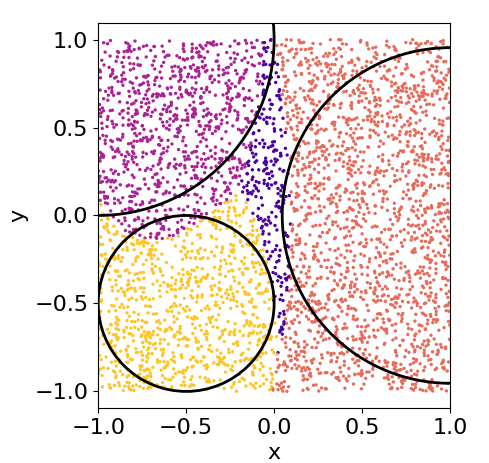}}
\subfigure[\hspace{0.05cm} 4 layers]{\includegraphics[width=0.22\textwidth]{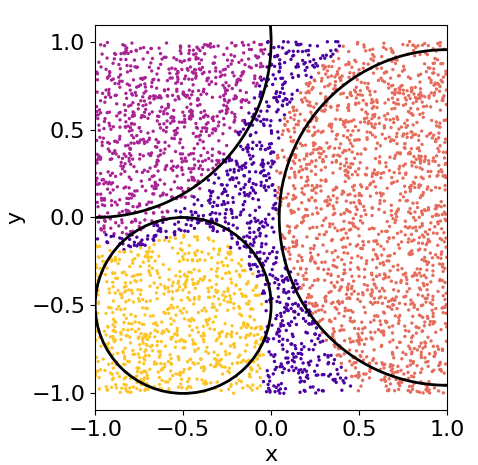}}
\subfigure[\hspace{0.05cm} 10 layers]{\includegraphics[width=0.22\textwidth]{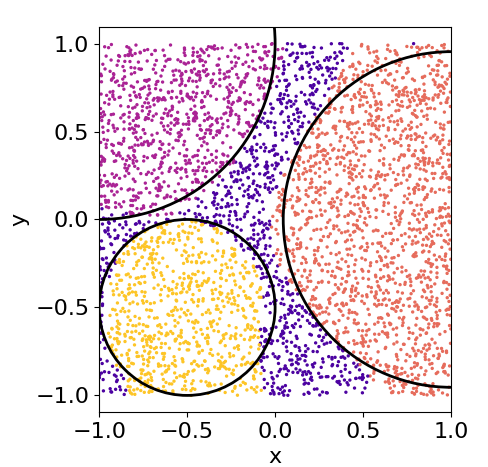}}
\caption{Results for the 3-circles problem using a single-qubit classifier trained with the L-BFGS-B minimizer and the weighted fidelity cost function. With one layer, the classifier intuits the four regions although the central one is difficult to tackle. With more layers, this region is clearer for the classifier and it tries to adjust the circular regions.}
\label{Fig:3circles_evol}
\end{figure}

As with the previous problem, it is interesting to compare the performance in terms of sucess rate of classifiers with different number of layers. Figure \ref{Fig:3circles_evol} shows the results for a two-qubit classifier with no entanglement for 1, 3, 4 and 10 layers. Even with only one layer, the classifier identifies the four regions, being the more complicated to describe the central one. As we consider more layers, the classifier performs better and adjust these four regions.

\subsection{Classification in multiple dimensions}

\begin{table*}[t!]
\centering
\begin{tabular}{c|c|cc|c|cc|cc}
  & \multicolumn{3}{c|}{$\chi^{2}_{f}$} & \multicolumn{5}{c}{$\chi^{2}_{wf}$} \\
 \hline
Qubits & 1 & \multicolumn{2}{c|}{2 } & 1 & \multicolumn{2}{c|}{2} & \multicolumn{2}{c}{4 }   \\
Layers  & & No Ent. & Ent. & & No Ent. & Ent. & No Ent. & Ent. \\ 
 \hline
 1 & 0.87 & 0.87 & -- & 0.87 & 0.87 & -- & 0.90 & -- \\
 2 & 0.87 & 0.87 & 0.87 & 0.87 & 0.92 & 0.91 & 0.90 & 0.98 \\
 3 & 0.87 & 0.87 & 0.87 & 0.89 & 0.89 & 0.97 & -- & -- \\
 4 & 0.89 & 0.87 & 0.87 & 0.90 & 0.93 & 0.97 & -- & -- \\
 5 & 0.89 & 0.87 & 0.87 & 0.90 & 0.93 & 0.98 & -- & -- \\
 6 & 0.90 & 0.87 & 0.87 & 0.95 & 0.93 & 0.97 & -- & -- \\
 8 & 0.91 & 0.87 & 0.87 & 0.97 & 0.94 & 0.97 & -- & -- \\
10 & 0.90 & 0.87 & 0.87 & 0.96 & 0.96 & 0.97 & -- & -- \\
\end{tabular}
\caption{Results of the single- and multi-qubit classifiers with data re-uploading for the four-dimensional hypersphere problem. Numbers indicate the success rate, i.e. the number of data points classified correctly over the total number of points. Words ``Ent." and ``No Ent." refer to considering entanglement between qubits or not, respectively. We have used the L-BFGS-B minimization method with the weighted fidelity and fidelity cost functions. The fidelity cost function gets stuck in some local minima for the multi-qubit classifiers. The results obtained with the weighted fidelity cost function are much better, reaching the 0.98 with only two layers for the four-qubit classifier. Here, the introduction of entanglement improves significantly the performance of the multi-qubit classifier.}
\label{tab:results_hypersphere}
\end{table*}

\begin{table*}[t!]
\centering
\begin{tabular}{c|c|cc|c|cc|cc}
  & \multicolumn{3}{c|}{$\chi^{2}_{f}$} & \multicolumn{5}{c}{$\chi^{2}_{wf}$} \\
 \hline
Qubits & 1 & \multicolumn{2}{c|}{2 } & 1 & \multicolumn{2}{c|}{2} & \multicolumn{2}{c}{4 }   \\
Layers  & & No Ent. & Ent. & & No Ent. & Ent. & No Ent. & Ent. \\ 
 \hline
 1 & 0.34 & 0.51 & -- & 0.43 & 0.77 & -- & 0.81 & -- \\
 2 & 0.57 & 0.63 & 0.59 & 0.76 & 0.79 & 0.82 & 0.87 & 0.96 \\
 3 & 0.80 & 0.68 & 0.65 & 0.68 & 0.94 & 0.95 & 0.92 & 0.94 \\
 4 & 0.84 & 0.78 & 0.89 & 0.79 & 0.93 & 0.96 & 0.93 & 0.96 \\
 5 & 0.92 & 0.86 & 0.82 & 0.88 & 0.96 & 0.96 & 0.96 & 0.95 \\
 6 & 0.93 & 0.91 & 0.93 & 0.91 & 0.93 & 0.96 & 0.97 & 0.96 \\
 8 & 0.90 & 0.89 & 0.90 & 0.92 & 0.94 & 0.95 & 0.95 & 0.94 \\
10 & 0.90 & 0.91 & 0.92 & 0.93 & 0.95 & 0.96 & 0.95 & 0.95 \\
\end{tabular}
\caption{Results of the single- and multi-qubit classifiers with data re-uploading for the three-class annulus problem. Numbers indicate the success rate, i.e. the number of data points classified correctly over the total number of points. Words ``Ent." and ``No Ent." refer to considering entanglement between qubits or not, respectively. We have used the L-BFGS-B minimization method with the weighted fidelity and fidelity cost functions. The weighted fidelity cost function presents better success rates than the fidelity cost function. The multi-qubit classifiers improve the results obtained with the single-qubit classifier but the using of entanglement does not introduce significant changes.}
\label{tab:results_3crown}
\end{table*}

\begin{figure*}[t!]
\centering
\subfigure[\hspace{0.05cm} 1 layer ]{\includegraphics[width=0.22\textwidth]{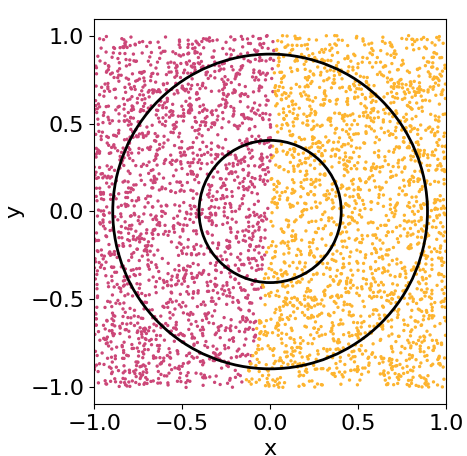}}
\subfigure[\hspace{0.05cm} 2 layers]{\includegraphics[width=0.22\textwidth]{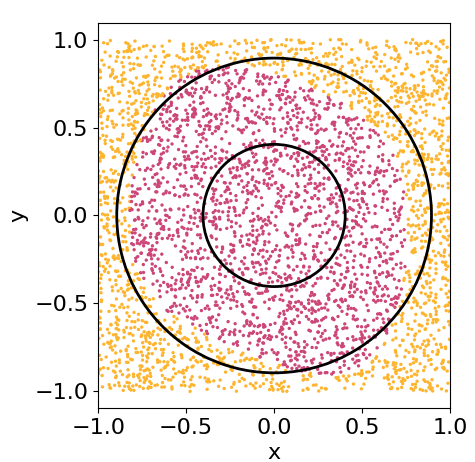}}
\subfigure[\hspace{0.05cm} 3 layers]{\includegraphics[width=0.22\textwidth]{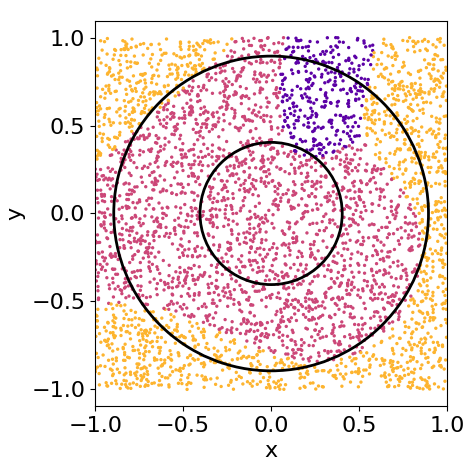}}
\subfigure[\hspace{0.05cm} 4 layers]{\includegraphics[width=0.22\textwidth]{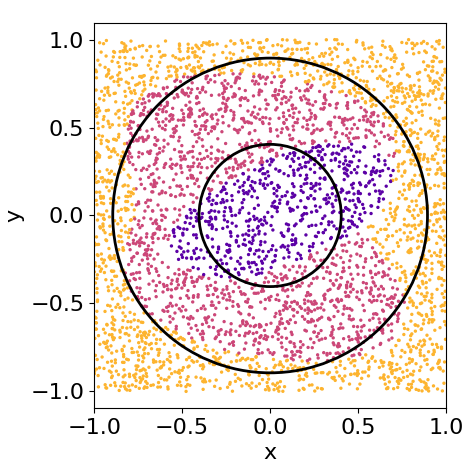}}
\subfigure[\hspace{0.05cm} 5 layers]{\includegraphics[width=0.22\textwidth]{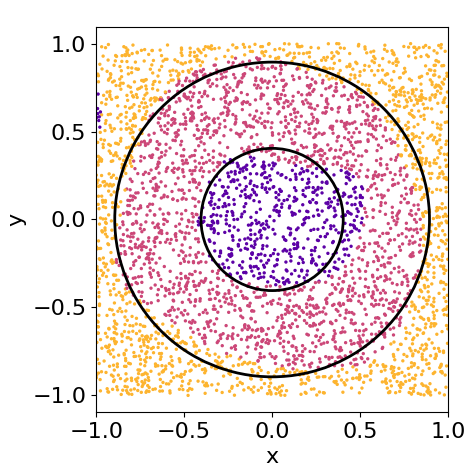}}
\subfigure[\hspace{0.05cm} 6 layers]{\includegraphics[width=0.22\textwidth]{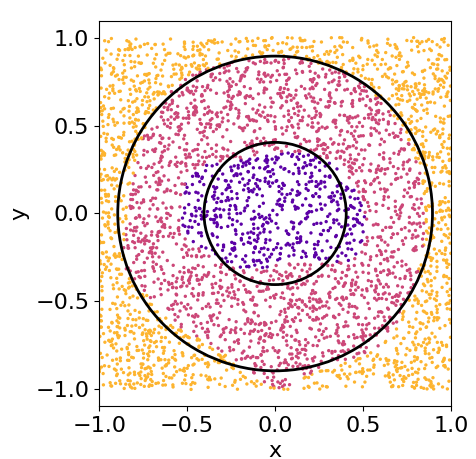}}
\subfigure[\hspace{0.05cm} 8 layers]{\includegraphics[width=0.22\textwidth]{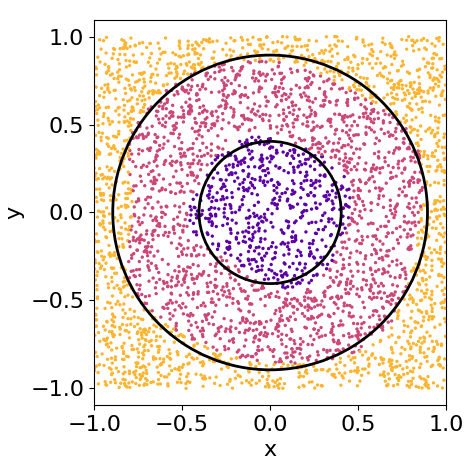}}
\subfigure[\hspace{0.05cm} 10 layers]{\includegraphics[width=0.22\textwidth]{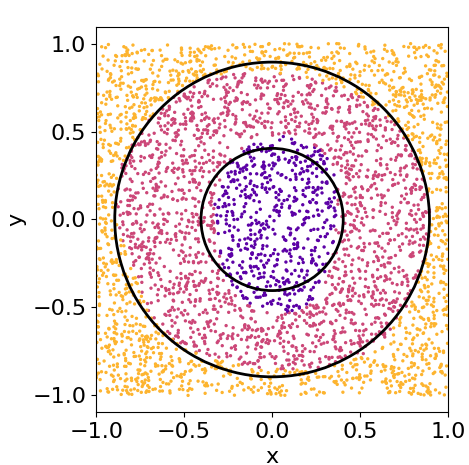}}

\caption{Results obtained with the single-qubit classifier for the annulus problem, using the weighted fidelity cost function during the training. The better results are obtained with a 10 layers classifier (93\% of success rate). As we consider more qubits and entanglement, we can increase the success rate up to 96\%, as shows Table \ref{tab:results_3crown}.}
\label{Fig:3crown_evol}
\end{figure*}

As explained in Section \ref{sec:structure}, there is no restriction in uploading multidimensional data. We can upload up to three values per rotation since this is the degrees of freedom of a SU(2) matrix. If the dimension of data is larger than that, we can just split the data vector into subsets and upload each one at a time, as described explicitly in Eq. \eqref{eq:multiple_dim}. Therefore, there is no reason to limit the dimension of data to the number of degrees of freedom of a qubit. We can in principle upload any kind of data if we apply enough gates. 

Following this idea we will now move to a more complicated classification using data with 4 coordinates. We use as a problem the four-dimensional sphere, i.e. classifying data points according to $x_{1}^2 + x_{2}^2 + x_{3}^2 + x_{4}^2< 2/\pi$. Similarly with the previous problems, $x_{i}\in[-1,1]$ and the radius has been chosen such that the volume of the hypersphere is half of the total volume. This time, we will take 1000 random points as the training set because the total volume increases. 

Results are shown in Table \ref{tab:results_hypersphere}. A single-qubit achieves 97\% of success with eight layers (82 parameters) using the weighted fidelity cost function. Results are better if we consider more qubits. For two qubits, the best result is 98\% and it only requires three entangled layers (62 parameters). For four qubits, it achieves 98\% success rate with two layers with entanglement, i.e. 82 parameters.

\subsection{Classification of non-convex figures}

As a final benchmark, we propose the classification of a non-convex pattern. In particular, we classify the points of an annulus with radii $r_{1}=\sqrt{0.8-2/\pi}$ and $r_{2}=\sqrt{0.8}$. We fix three classes: points inside the small circle, points in the annulus and points outside the big circle. So, besides it being a non-convex classification task, it is also a multi-class problem. A simpler example, with binary classification, can be found in Appendix \ref{app:results}.

The results are shown in Table \ref{tab:results_3crown}. It achieves 93\% of success with a single-qubit classifier with 10 layers and a weighted fidelity cost function. With two qubits, it achieves better results, 94\% with three layers. With four qubits, it reaches a 96\% success rate with only two layers with entanglement.

It is interesting to observe how the single-qubit classifier attempts to achieve the maximum possible results as we consider more and more layers. Figure \ref{Fig:3crown_evol} shows this evolution in terms of the number of layers for a single-qubit classifier trained with the weighted fidelity cost function. It requires four layers to learn that there are three concentric patterns and the addition of more layers adjusts these three regions.

\subsection{Comparison with classical classifiers}

It is important to check if our proposal is in some sense able to compete with actual technology of supervised machine learning. To do so we have used the standard machine learning library \texttt{scikit-learn} \citep{scikit-learn} and solved the same problems as we have solved with the quantum classifier. We have included the four problems presented in the main paper plus five extra problems analyzed in Appendix \ref{app:results}. The aim of this classical benchmarking is not to make an extended review of what classical machine learning is capable to perform. The aim is to compare our simple quantum classifier to simple models such as shallow neural networks and simple support vector machines. 

The technical details of the classical classification are the following: the neural network has got one hidden layer with 100 neurons, a ReLu activation function and the solver \emph{lbfgs} by \texttt{scikit-learn}. The support vector machine is the default \texttt{sklearn.svm.SVC}. Some changes in the initialization parameters were tested with no significant differences.

Table \ref{tab:classical_benchmark} compares the best performance of a neural network, support vector classifier (SVC), the single-qubit classifier with fidelity cost function and single-qubit classifier with a weighted fidelity cost function.
In all problems, the performance of the single-qubit classifier is, at least, comparable with the classical methods. In some problems, like the 3-circles problem and the binary annulus problem, the results of the single-qubit classifier are better than with the classical methods. 

\begin{table*}[t!]
\centering
\begin{tabular}{c|cc|cc}
\multirow{2}{*}{Problem} & \multicolumn{2}{c|}{Classical classifiers } & \multicolumn{2}{c}{Quantum classifier } \\
 & NN & SVC & $\chi_{f}^2$ & $\chi^2_{wf}$ \\ \hline
 Circle & 0.96 & 0.97 & 0.96 & 0.97 \\
 3 circles & 0.88 & 0.66 & 0.91 & 0.91 \\
 Hypersphere & 0.98 & 0.95 & 0.91 & 0.98 \\
 Annulus & 0.96 & 0.77 & 0.93 & 0.97 \\
 Non-Convex & 0.99 & 0.77 & 0.96 & 0.98 \\
 Binary annulus & 0.94 & 0.79 & 0.95 & 0.97 \\
 Sphere & 0.97 & 0.95 & 0.93 & 0.96 \\
 Squares & 0.98 & 0.96 & 0.99 & 0.95 \\
 Wavy Lines & 0.95 & 0.82 & 0.93 & 0.94 \\ 
\end{tabular}
\caption{Comparison between single-qubit quantum classifier and two well-known classical classification techniques: a neural network (NN) with a single hidden layer composed of 100 neurons and a support vector classifier (SVC), both with the default parameters as defined in \texttt{scikit-learn} python package. We analyze nine problems: the first four are presented in Section \ref{sec:benchmark} and the remaining five in Appendix \ref{app:results}.
Results of the single-qubit quantum classifier are obtained with the fidelity and weighted fidelity cost functions, $\chi^2_{f}$ and $\chi^{2}_{wf}$ defined in Eq. \eqref{eq:fidelity_chi2} and Eq. \eqref{eq:conventional_chi2} respectively. This table shows the best success rate, being 1 the perfect classification, obtained after running ten times the NN and SVC algorithms and the best results obtained with single-qubit classifiers up to 10 layers.}
\label{tab:classical_benchmark}
\end{table*}

\section{Conclusions \label{sec:conclusions}}

We have proposed a single-qubit classifier that can represent multidimensional complex figures. The core of this quantum classifier is the data re-uploading. This formalism allows circumventing the limitations of the no-cloning theorem to achieve a correct generalization of an artificial neural network with a single layer. In that sense, we have applied the Universal Approximation Theorem to prove the universality of a single-qubit classifier.

The structure of this classifier is the following. Data and processing parameters are uploaded multiple times  along the circuit by using one-qubit rotations. The processing parameters of these rotations are different at each upload and should be optimized using a classical minimization algorithm. To do so, we have defined two cost functions: one inspired in the traditional neural networks cost functions (weighted fidelity cost function) and the other, simpler, consisting of the computation of the fidelity of the final state with respect to a target state. These target states are defined to be maximally orthogonal among themselves. Then, the single-qubit classifier finds the optimal rotations to separate the data points into different regions of the Bloch sphere, each one corresponding with a particular class.

The single-qubit classifier can be generalized to a larger number of qubits. This allows the introduction of entanglement between these qubits by adding two-qubit gates between each layer of rotations. We use a particular entangling ansantz as a proof of concept. The exploration of other possible ansatzes is out of the scope of this work.

We have benchmarked several quantum classifiers of this kind, made of a different number of layers, qubits and with and without entanglement. The patterns chosen to test these classifiers are the points inside and outside of a circle (simple example) and similarly for a four-dimensional hypersphere (multidimensional example); a two dimensional region composed by three circles of different size (multiple classes example); and the points outside and inside of an annulus (non-convex example). In all cases, the single-qubit classifier achieves more than 90\% of the success rate. The introduction of more qubits and entanglement increases this success and reduces the number of layers required. The weighted fidelity cost function turns out to be more convenient to achieve better results than the fidelity cost function. In all problems, the probability to get stuck in a local minima increases with the number of layers, an expected result from an optimization problem involving several parameters.

In summary, we have proposed a quantum classifier model that seems to be universal by exploiting the non-linearities of the single-qubit rotational gates and by re-uploading data several times.

\section*{Acknowledgements}
This work has been supported by Project Quantum CAT (001-P-001644). APS and JIL acknowledge CaixaBank for its support of this work through Barcelona Supercomputing Center project \emph{CaixaBank Computaci\'{o}n Cu\'{a}ntica}. The authors acknowledge the interesting discussions with Quantic group team members. The authors also acknowledge Xanadu Quantum Computing, in particular Shahnawaz Ahmed, for writing a tutorial of this quantum classifier using its full-stack library Pennylane \cite{pennylane}.

\bibliographystyle{apsrev4-1}
\bibliography{Citations}

\appendix
\section{The Stochastic Gradient Descent method (SGD)\label{sec:SGD}}

We propose an algorithm to compute the derivatives needed to minimize the cost function that is inspired in the stochastic gradient descent algorithm for neural networks. This kind of computation is not new, as similar proposals can be found in \cite{SBSW18,FN18,RBMHLA18}. We show that, in analogy to neural networks, this algorithm can be interpreted as a back-propagation, as it takes intermediate steps of the circuit for exactly computing the gradient of the function $\chi^2_{f}$. However, it requires full access to the wave function at intermediate steps of the computation, which makes it costly to be implemented experimentally.

\subsection{To compute the gradient}

Our goal is to find the gradient of $\chi^2_{f}$ for one data point of the training set. In terms of Eq. \eqref{eq:qcircuit} and Eq. \eqref{eq:layers},
\begin{equation}
|\braket{\tilde{\psi}_{c}}{\psi}|^2 = |\langle\tilde{\psi}_{c}|L(N)L(N-1)\cdots L(1)|0\rangle|^2.
\label{eq:chi2_SGD}
\end{equation} 
Let us write explicitly an intermediate quantum state $|\psi_{l}\rangle$ of the single-qubit classifier,
\begin{equation}
\ket{\psi_l} = \prod_{k = 0}^l L(k) \ket 0 = L(l)\ket{\psi_{l-1}},
\end{equation}
where $1\leq l\leq N$ and $\ket{\psi_{0}}\equiv|0\rangle$. Next, notice that we can recursively define the counterpart of the label states $|\tilde{\psi}_{c}\rangle$ as 
\begin{equation}
\bra{\Delta_{l}}=\bra{\tilde{\psi}_{c}} L(N) L(N-1) \ldots L(l+1) = \bra{\Delta_{l+1}}L(l+1),
\end{equation}
where $1\leq l\leq N$ and $\bra{\Delta_{N}}\equiv\bra{\tilde{\psi}_{c}}$. A way to understand this notation is by checking that $\ket{\psi_l}$ counts gates from the beginning of the circuit, starting from first layer $L(1)$, and $\bra{\Delta_{l}}$ counts backwards from the end of the circuit, starting from $N^{\mathrm{th}}$ layer. 

Thus, Eq. \eqref{eq:chi2_SGD} can be rewritten as
\begin{equation}
|\braket{\tilde{\psi}_{c}}{\psi}|^2 = |\braket{\Delta_{l}}{\psi_l}|^2,
\end{equation}
where $l\in[1,N]$ defines the circuit layer where the gradient will be computed.

Each layer $L(l)$ contains as much as six tunable parameters: three rotational angles represented with the vector $\vec{\theta}_{l}$ and three weights $\vec{w}_{l}$. For a problem of dimension $d>3$, each layer is composed by sublayers with up to six tunable parameters, as shown in Eq.\eqref{eq:multiple_dim}. Then, when we derive with respect to one parameter, only the corresponding layer gate will be affected by it. We can split the set of gates in two parts: those before the derivative gate and those after. Using the recursive definitions presented above, the derivative with respect to $\vec{\theta}_{l}=(\theta^{1}_{l},\theta^{2}_{l},\theta^{3}_{l})$ angles can be written as
\begin{equation}
\frac{\partial \chi_{f}^2}{\partial\theta^{i}_l} = - 2 Re\left\lbrace\left(\,\bra{\Delta_{l}}\frac{\partial L(l)}{\partial\theta^{i}_l}\ket{\psi_{l-1}}\right)\,\braket{\psi_l}{\Delta_{l}}\right\rbrace ,
\end{equation}
for $i=1,2,3$ and similarly with $\partial \chi_{f}^2/\partial\vec{w}_l$. This derivative is not unitary. Nevertheless, this fact has not importance in our algorithm since we will use its value to update the values of the new parameters in the next iteration classically.

The computation of the partial derivatives of $L(l)$ is straightforward do to its matrix unitary structure. In general, $L(l)$ is a general SU(2) matrix which can be parametrized as
\begin{equation}
L(l)\equiv U(\vec{\phi}_{l}) = \begin{pmatrix}
\cos\frac{\phi^{1}_{l}}{2} e^{i\frac{(\phi^{2}_{l}+\phi^{3}_{l})}{2}}& -\sin\frac{\phi^{1}_{l}}{2} e^{-i\frac{(\phi^{2}_{l}-\phi^{3}_{l})}{2}} \\
\sin\frac{\phi^{1}_{l}}{2} e^{i\frac{(\phi^{2}_{l}-\phi^{3}_{l})}{2}} & \cos\frac{\phi^{1}_{l}}{2} e^{-i\frac{(\phi^{2}_{l}+\phi^{3}_{l})}{2}}
\end{pmatrix},
\label{eq:Uparam}
\end{equation}
where $\vec{\phi}_{l} =(\phi^{1}_{l},\phi^{2}_{l},\phi^{3}_{l})= \vec{\theta}_{l} + \vec{w}_{l}\circ\vec{x}$. Thus,
\begin{align}
\frac{\partial L(l)}{\partial \theta^{i}_{l}} &= \frac{\partial L(l)}{\partial \phi^{i}_{l}}, \\ 
\frac{\partial L(l)}{\partial w^{i}_{l}} &= \frac{\partial L(l)}{\partial \theta^{i}_{l}}x^{i},
\end{align}

The key feature of taking the $L(l)$ parametrization of Eq.\eqref{eq:Uparam} is that its derivatives are almost itself, which simplifies its computation significantly. Notice that 
\begin{equation}
\frac{\partial L(l)}{\partial \theta^i_{l}} = \frac{\partial U(\vec{\theta}_{l}+w_{l}\circ\vec{x})}{\partial \theta^i_{l}}=\frac{1}{2}U(\vec{\theta}_{l}+\vec{w}_{l}\circ\vec{x} + \pi\vec{\delta}_{i}),
\end{equation}
where $\vec{\delta}_{i}$ is a vector of dimension three which components are all zeros except the $i$-th one. The calculations for the derivative become simpler in this case, as we just have to shift the proper parameter by $\pi$, and use it to measure the derivative. 

Once we have computed the gradient of the cost function $\chi_{f}^2$, we can update the parameters by setting
\begin{align}
\theta^{i}_{l} &\leftarrow \theta^{i}_{l} - \eta \frac{\partial \chi_{f}^2}{\partial \theta^{i}_{l}}, \\
w^{i}_{l} &\leftarrow w^{i}_{l} - \eta \frac{\partial \chi_{f}^2}{\partial w^{i}_{l}}, 
\end{align}
where $\eta$ is an adjustable learning rate. \\

\subsection{To update parameters}

There are several ways of updating the parameters. One naive option is to update parameters every time we compute the gradient of the function. Then we start all over again and repeat until the parameters become stable. This is not a good practice. If we do so, we are moving point by point, optimizing and classifying every data from the training set of our classifier. This classifies all points slowly. At the late steps of the minimization process, we will have to face plenty of points which are already classified and do not have important gradients. In summary, this is a very inefficient algorithm.

Another option is to compute the gradients for every point, and then take the average of all gradients and update parameters with it. However, this method is likely to achieve poor accuracies. Some gradients can be canceled out by some others and the stability is achieved before it should be. 

The method that is usually used is the so-called \textit{batched optimization}. This method is a mix of the two previous ways of tackling the problem. First, we have to split the training set into smaller subsets or batches. We compute the averaged gradient of the first batch with respect to all points in the batch and then update. Then, proceed similarly with the second batch. We keep going until we finish all the batches. The crucial step is to shuffle the training set and split it into new different mixed batches. This way, we optimize the classifier respect to subsets of the original training set that are constantly changing which is more efficient than a point-by-point optimization, but statistically equivalent. In our SGD algorithm, we took batches with 20 points out of a total amount of 200 points for the two-dimensional problems, 500 for three-dimensional problems and 1000 for four-dimensional problems.

Although we tested this SGD algorithm, we obtained worse results than with the L-BFGS-B method. In addition, the computational efforts for this SGD algorithm were larger as could not be fully optimized. Thus we discard for the moment this SGD minimization method.

\section{Classification results in other problems \label{app:results}}

We tested our classifier in several different problems. Here, we show a summary of the results obtained in the other problems we tackled. As the ones defined in the main paper, we constrained the data space to $\vec{x} \in [-1,1]^{dim}$. These extra classification tasks aim to cover different kinds of training data, and show that our quantum classifier can adapt itself to large varieties of problems.

The datasets have 200 random entries -- except for the sphere, which has 500 -- and test the performance of the classifier against 4000 random points. All problems were designed without biases, i.e. a blind classifier will get successes of $\sim$50\% for a binary classifier, $\sim$33\% for a ternary classifier and $\sim$25\% for a quaternary classifier. 

In general, we can observe that the classifier improves its performance as we increase the number of layers until it comes into a stationary regime. More qubits and entanglement usually allows entering in the stationary regime with a lesser number of layers. 

In the following subsections, we define these extra problems and present the results of each one summarized in a table. The best results obtained in each problem are plotted in Figure \ref{Fig:Results_extra}.

\subsubsection*{Non-convex problem}

\begin{table*}[t!]
\centering
\begin{tabular}{c|c|cc|c|cc|cc}
  & \multicolumn{3}{c|}{$\chi^{2}_{f}$} & \multicolumn{5}{c}{$\chi^{2}_{wf}$} \\
 \hline
Qubits & 1 & \multicolumn{2}{c|}{2 } & 1 & \multicolumn{2}{c|}{2} & \multicolumn{2}{c}{4 }   \\
Layers  & & No Ent. & Ent. & & No Ent. & Ent. & No Ent. & Ent. \\ 
 \hline
 1 & 0.49 & 0.55 & -- & 0.49 & 0.76 & -- & 0.76 & -- \\
 2 & 0.82 & 0.75 & 0.75 & 0.86 & 0.94 & 0.85 & 0.96 & 0.96 \\
 3 & 0.93 & 0.74 & 0.85 & 0.96 & 0.95 & 0.95 & 0.95 & 0.97 \\
 4 & 0.93 & 0.74 & 0.88 & 0.95 & 0.96 & 0.97 & 0.95 & 0.96 \\
 5 & 0.91 & 0.95 & 0.90 & 0.97 & 0.95 & 0.96 & 0.95 & 0.97 \\
 6 & 0.96 & 0.94 & 0.93 & 0.98 & 0.97 & 0.97 & 0.95 & 0.97 \\
 8 & 0.96 & 0.96 & 0.95 & 0.98 & 0.98 & 0.97 & 0.96 & 0.97 \\
10 & 0.95 & 0.92 & 0.95 & 0.96 & 0.96 & 0.96 & 0.96 & 0.97 \\
\end{tabular}
\caption{Results of the single- and multi-qubit classifiers with data re-uploading for the non-convex problem. Numbers indicate the success rate, i.e. the number of data points classified correctly over the total number of points. Words ``Ent." and ``No Ent." refer to considering entanglement between qubits or not respectively. We have used the L-BFGS-B minimization method with the weighted fidelity and fidelity cost functions. Both cost functions lead to higher success rates, although the weighted fidelity cost function is better. It achieves the 0.98 success with two qubits, entanglement, and four layers.}
\label{tab:results_non_convex}
\end{table*}

The \emph{non-convex} problem is made for testing two mutually non-convex zones dividing a 2D area. The border of both zones lies on the function $x_2 = -2 x_1 + 3/2 \sin(\pi x_1)$.  This border is chosen in a way that there is no area small enough to leave it unclassified, forcing the classifier to catch these smaller zones. 

Complete results can be read in Table \ref{tab:results_non_convex}. The weighted fidelity cost function gets the best performance, 98\%, with 6 layers and 1 qubit, (32 parameters). Fidelity cost gets a result of 96\% for the same conditions, while the best, 97\%, is achieved for 2 entangled qubits with 8 layers (80 parameters). The final figure obtained after the classification can be seen in Figure \ref{Fig:non convex}.

We can compare this result with the one obtained in Table \ref{tab:classical_benchmark}. Neural networks work well as they can approximate our non-convex function properly. However, SVC performs worse, as they find hard to deal with non-convexity. In contrast with that, our quantum classifier is plastic enough to classify this data with no major difficulty.

\subsubsection*{Binary annulus}

\begin{table*}[t!]
\centering
\begin{tabular}{c|c|cc|c|cc|cc}
  & \multicolumn{3}{c|}{$\chi^{2}_{f}$} & \multicolumn{5}{c}{$\chi^{2}_{wf}$} \\
 \hline
Qubits & 1 & \multicolumn{2}{c|}{2 } & 1 & \multicolumn{2}{c|}{2} & \multicolumn{2}{c}{4 }   \\
Layers  & & No Ent. & Ent. & & No Ent. & Ent. & No Ent. & Ent. \\ 
 \hline
 1 & 0.44 & 0.50 & -- & 0.44 & 0.59 & -- & 0.66 & -- \\
 2 & 0.48 & 0.50 & 0.51 & 0.53 & 0.73 & 0.72 & 0.70 & 0.96 \\
 3 & 0.91 & 0.50 & 0.56 & 0.74 & 0.75 & 0.95 & 0.78 & 0.96 \\
 4 & 0.80 & 0.74 & 0.56 & 0.86 & 0.97 & 0.97 & 0.92 & 0.96 \\
 5 & 0.90 & 0.93 & 0.88 & 0.89 & 0.97 & 0.96 & 0.97 & 0.94 \\
 6 & 0.92 & 0.91 & 0.94 & 0.95 & 0.94 & 0.95 & 0.95 & 0.93 \\
 8 & 0.90 & 0.93 & 0.95 & 0.92 & 0.94 & 0.94 & 0.96 & 0.94 \\
10 & 0.90 & 0.92 & 0.91 & 0.92 & 0.95 & 0.93 & 0.96 & 0.93 \\
\end{tabular}
\caption{Results of the single- and multi-qubit classifiers with data re-uploading for the binary annulus problem. Numbers indicate the success rate, i.e. the number of data points classified correctly over the total number of points. Words ``Ent." and ``No Ent." refer to considering entanglement between qubits or not respectively. We have used the L-BFGS-B minimization method with the weighted fidelity and fidelity cost functions. As happens in other problems, the results obtained with the weighted fidelity cost function are better than the ones obtained with the fidelity cost function. The multi-qubit classifiers and the introduction of entanglement increase the success rates.}
\label{tab:results_binary_annulus}
\end{table*}

This a binary version of the annulus problem. The two classes are defined as being inside or outside the annulus with radii $r_1 = \sqrt{0.8 - 2/\pi}$ and $r_2 = \sqrt{0.8}$. 
This geometry, as happens with the three-class annulus problem, is interesting because the classifier finds a way to connect areas of the problem which are disconnected. This involves a to and fro path for the parameters. 

The fidelity cost function reaches the best result at 94\%, with 2 qubits with entanglement and 6 layers, which is 60 parameters. For the weighted fidelity, we reach up to 97\% with 2 qubits with no entanglement and 4 layers, 40 parameters. The results are written in Table \ref{tab:results_binary_annulus} and the best result is plotted in Figure \ref{Fig:crown}. 

If we compare these results to the classical ones in Table \ref{tab:classical_benchmark}, we see again that the SVC method is way worse than the NN, which can be interpreted as the sign of non-convexity this problem has.

\subsubsection*{Sphere}

\begin{table*}[t!]
\centering
\begin{tabular}{c|c|cc|c|cc|cc}
  & \multicolumn{3}{c|}{$\chi^{2}_{f}$} & \multicolumn{5}{c}{$\chi^{2}_{wf}$} \\
 \hline
Qubits & 1 & \multicolumn{2}{c|}{2 } & 1 & \multicolumn{2}{c|}{2} & \multicolumn{2}{c}{4 }   \\
Layers  & & No Ent. & Ent. & & No Ent. & Ent. & No Ent. & Ent. \\ 
 \hline
 1 & 0.53 & 0.70 & -- & 0.53 & 0.70 & -- & 0.70 & --\\
 2 & 0.77 & 0.73 & 0.53 & 0.78 & 0.94 & 0.96 & 0.96 & 0.96 \\
 3 & 0.76 & 0.74 & 0.77 & 0.78 & 0.92 & 0.94 & 0.94 & 0.95 \\
 4 & 0.84 & 0.83 & 0.78 & 0.89 & 0.92 & 0.94 & 0.95 & 0.94 \\
 5 & 0.89 & 0.85 & 0.77 & 0.90 & 0.94 & 0.94 & 0.95 & 0.94 \\
 6 & 0.90 & 0.89 & 0.88 & 0.92 & 0.87 & 0.93 & 0.94 & 0.94 \\
 8 & 0.89 & 0.87 & 0.90 & 0.93 & 0.92 & 0.89 & 0.94 & 0.93 \\
10 & 0.93 & 0.91 & 0.90 & 0.93 & 0.94 & 0.92 & 0.92 & 0.92 \\
\end{tabular}
\caption{Results of the single- and multi-qubit classifiers with data re-uploading for the three-dimensional sphere problem. Numbers indicate the success rate, i.e. the number of data points classified correctly over the total number of points. Words ``Ent." and ``No Ent." refer to considering entanglement between qubits or not respectively. We have used the L-BFGS-B minimization method with the weighted fidelity and fidelity cost functions. The weighted fidelity cost function is better than the fidelity cost function. There are no significant differences between the two-qubit and the four-qubit classifiers. Both are better than the single-qubit classifier and the introduction of entanglement does not increase the success rates.}
\label{tab:results_sphere}
\end{table*}

This quantum classifier is able to classify multidimensional data, as we have shown with the four-dimensional hypersphere. We also tested a three-dimensional figure, a sphere with radius $r = \sqrt[3]{3/\pi}$.

For this problem, the fidelity cost function reaches its maximum, 93\%, with a single-qubit classifier of 10 layers (60 parameters). The same success is obtained with a two-qubit entangled classifier and 6 layers (72 parameters). With the weighted fidelity, this success rate grows up to 96\% for two- and four- qubit classifier of 2 layers (24 and 48 parameters respectively) with and without entanglement. 
All results are written in Table \ref{tab:results_sphere}.

\subsubsection*{Squares}

This problem divides a 2D area into four quadrants with straight lines. This is one of the easiest problems for a neural network. By construction, neural networks can establish a separation between classes by using biases, and thus straight lines are immediate to understand. We construct this problem to see how a quantum classifier performs against a neural network in the latter's field. 

\begin{figure*}[t!]
\centering
\subfigure[\hspace{0.05cm} $\chi^2_{wf}$, 1 qubit, 6 layers \label{Fig:non convex}]{\includegraphics[width=0.45\textwidth]{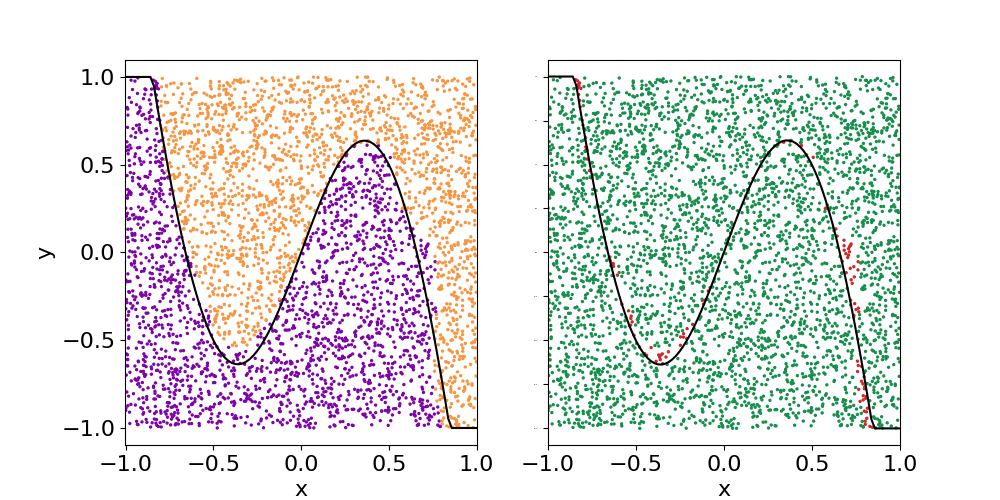}}
\subfigure[\hspace{0.05cm} $\chi^2_{wf}$, 2 qubits without entanglement, 4 layers \label{Fig:crown}]{\includegraphics[width=0.45\textwidth]{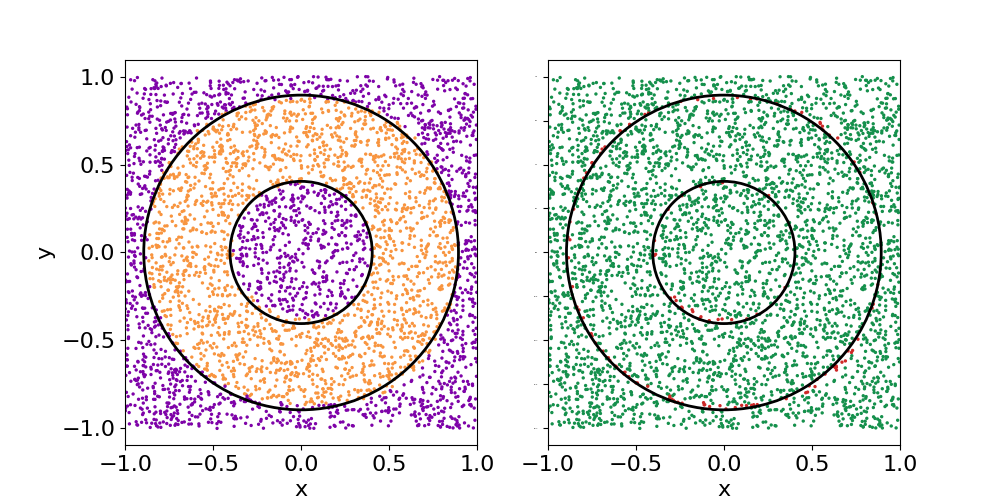}}
\subfigure[\hspace{0.05cm} $\chi^2_{f}$, 2 qubits without entanglement, 6 layers \label{Fig:squares}]{\includegraphics[width=0.45\textwidth]{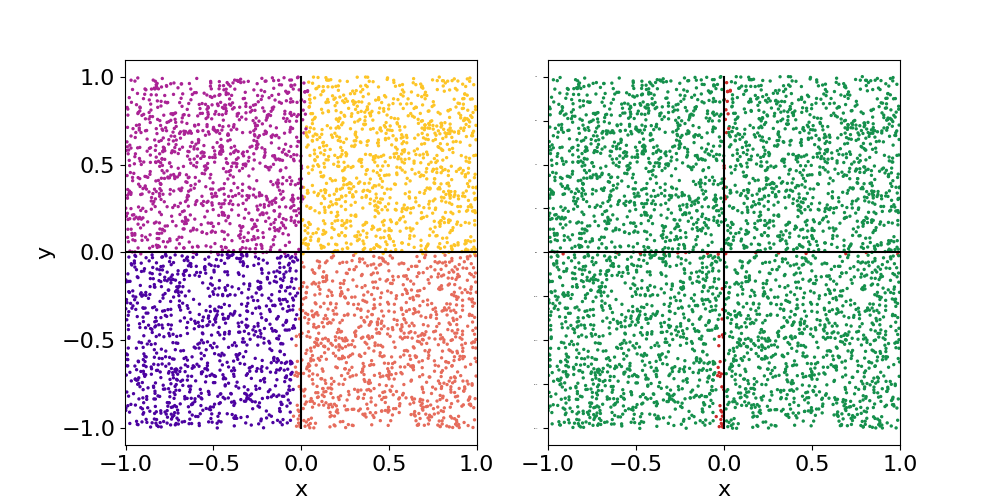}}
\subfigure[\hspace{0.05cm} $\chi^2_{wf}$, 2 qubits with entanglement, 6 layers \label{Fig:wavy}]{\includegraphics[width=0.45\textwidth]{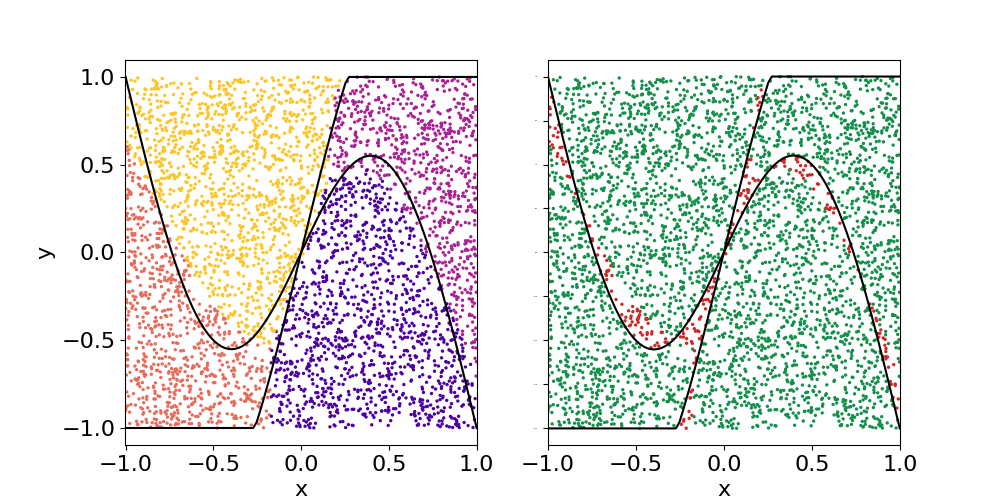}}
\caption{The best results for the 2D problems analyzed in this appendix. The caption of each figure includes the architecture of the classifier, i.e. number of qubits, layers and the use of entanglement. In the case of equal results with different classifiers, the simplest architecture has been chosen. Colors in the left part of each figure represent the different classes obtained from the classifier outputs. The right part of each figure prints the points correctly classified (in green) and bad classified (in red). Black solid lines define the problem boundaries. Notice that all missed points are located near the borders of the problem. This means that the classifier is understanding properly which is the problem most important features, but it lacks more training points. This can be easily corrected if we increase the number of training points.}
\label{Fig:Results_extra}
\end{figure*}

Classical results are very good, up to 98\% and 96\% for neural networks of 100 neurons and a single hidden layer and SVC respectively. However, quantum classifier performs even better. The fidelity cost function reaches 99\% of success in a two-qubit classifier without entanglement and 6 layers (60 parameters). Any two-qubit result is comparable with the success rate of the classical models. Something similar can be found for the weighted fidelity. The maximum success, 96\%, is obtained with a two-qubit entangled classifier with 4 layers (40 parameters). The results are written in Table \ref{tab:results_squares} and the best performance is plotted in Figure \ref{Fig:squares}.

\begin{table*}[t!]
\centering
\begin{tabular}{c|c|cc|c|cc|cc}
  & \multicolumn{3}{c|}{$\chi^{2}_{f}$} & \multicolumn{5}{c}{$\chi^{2}_{wf}$} \\
 \hline
Qubits & 1 & \multicolumn{2}{c|}{2 } & 1 & \multicolumn{2}{c|}{2} & \multicolumn{2}{c}{4 }   \\
Layers  & & No Ent. & Ent. & & No Ent. & Ent. & No Ent. & Ent. \\ 
 \hline
 1 & 0.58 & 0.48 & -- & 0.70 & 0.92 & -- & 0.90 & -- \\
 2 & 0.76 & 0.96 & 0.97 & 0.74 & 0.91 & 0.94 & 0.95 & 0.95 \\
 3 & 0.90 & 0.96 & 0.98 & 0.90 & 0.94 & 0.95 & 0.95 & 0.95 \\
 4 & 0.89 & 0.98 & 0.96 & 0.88 & 0.94 & 0.95 & 0.95 & 0.95 \\
 5 & 0.91 & 0.97 & 0.98 & 0.89 & 0.94 & 0.94 & 0.95 & 0.94 \\
 6 & 0.92 & 0.99 & 0.94 & 0.93 & 0.94 & 0.94 & 0.94 & 0.94 \\
 8 & 0.93 & 0.98 & 0.94 & 0.93 & 0.94 & 0.95 & 0.95 & 0.94 \\
10 & 0.94 & 0.97 & 0.93 & 0.94 & 0.94 & 0.94 & 0.94 & 0.93 \\
\end{tabular}
\caption{Results of the single- and multi-qubit classifiers with re-uploading data for the four-classes squares problem. Numbers indicate the success rate, i.e. the number of data points classified correctly over the total number of points. Words ``Ent." and ``No Ent." refer to considering entanglement between qubits or not respectively. We have used the L-BFGS-B minimization method with the weighted fidelity and fidelity cost functions. In this problem, the fidelity cost function presents better results than the weighted fidelity cost function. It achieves the 0.99 success with the two-qubit classifier with six layers and no entanglement.}
\label{tab:results_squares}
\end{table*}

\begin{table*}[t!]
\centering
\begin{tabular}{c|c|cc|c|cc|cc}
  & \multicolumn{3}{c|}{$\chi^{2}_{f}$} & \multicolumn{5}{c}{$\chi^{2}_{wf}$} \\
 \hline
Qubits & 1 & \multicolumn{2}{c|}{2 } & 1 & \multicolumn{2}{c|}{2} & \multicolumn{2}{c}{4 }   \\
Layers  & & No Ent. & Ent. & & No Ent. & Ent. & No Ent. & Ent. \\ 
 \hline
 1 & 0.70 & 0.52 & -- & 0.76 & 0.75 & -- & 0.88 & -- \\
 2 & 0.86 & 0.75 & 0.80 & 0.84 & 0.89 & 0.88 & 0.91 & 0.92 \\
 3 & 0.74 & 0.82 & 0.84 & 0.84 & 0.92 & 0.91 & 0.92 & 0.92 \\
 4 & 0.80 & 0.85 & 0.87 & 0.87 & 0.89 & 0.93 & 0.92 & 0.93 \\
 5 & 0.85 & 0.90 & 0.88 & 0.87 & 0.92 & 0.92 & 0.93 & 0.93 \\
 6 & 0.92 & 0.92 & 0.91 & 0.88 & 0.93 & 0.94 & 0.93 & 0.93 \\
 8 & 0.90 & 0.91 & 0.91 & 0.92 & 0.92 & 0.92 & 0.93 & 0.94 \\
10 & 0.92 & 0.91 & 0.93 & 0.90 & 0.93 & 0.93 & 0.93 & 0.93 \\
\end{tabular}
\caption{Results of the single- and multi-qubit classifiers with re-uploading data for the four-classes wavy lines problem. Numbers indicate the success rate, i.e. the number of data points classified correctly over the total number of points. Words ``Ent." and ``No Ent." refer to considering entanglement between qubits or not respectively. We have used the L-BFGS-B minimization method with the weighted fidelity and fidelity cost functions. Results with the weighted fidelity cost function are slightly better than the ones obtained with the fidelity cost function. The multi-qubit classifiers are vaguely better than the single-qubit classifiers. The introduction of entanglement does not change significantly the results.}
\label{tab:results_wavy}
\end{table*}

\subsubsection*{Wavy lines}

This problem is the four-class version of the non-convex problem. Now the area is divided into four regions by two different functions. The borders' equations are $x_2 =  \sin(\pi x_1) \pm x_1$. The important feature of this problem is that there are some areas in the problem too small to be caught by the classifier. 

As can be seen in Figure \ref{Fig:wavy}, most of the failure points are in these small non-convex areas. The classifier would rather adjust the rest of the points instead of tuning those zones and losing everything else. The results for this problem are not as good as for other problems, but we still get 94\% for the fidelity cost function, two entangled qubits and 10 layers (200 parameters) and the weighted fidelity, four entangled qubits and 4 layers (80 parameters). 

It is remarkable to compare these results, written in Table \ref{tab:results_wavy}, with the ones obtained using classical models. The quantum classifier approximately equals the NN method for this problem and outperforms SVC, 94\% against 95\% (NN) and 82\% (SVC). 

\end{document}